\documentclass[11pt]{amsart}
\usepackage{amsbsy,amssymb,amscd,amsfonts,latexsym,amstext,delarray,
amsmath,graphicx} 

\newtheorem{thm}{Theorem}[section]
\newtheorem{prop}[thm]{Proposition}

\newtheorem{lem}[thm]{Lemma}

\newtheorem{rem}[thm]{Remark}

\numberwithin{equation}{section}

\def\Z{{\mathbb Z}}
\def\R{{\mathbb R}}

\def\cB{{\mathcal B}}

\def\cP{{\mathcal P}}

\def\cR{{\mathcal R}}

\def\cV{{\mathcal V}}
\def\cW{{\mathcal W}}

\def\Tr{{\rm Tr}}

\title[Coupling topology and inflation in NCG cosmology]{The coupling of topology and inflation in noncommutative cosmology}
\author{Matilde Marcolli}
\author{Elena Pierpaoli}
\author{Kevin Teh}
\address{Department of Mathematics  \\
California Institute of Technology \\ 
Pasadena, CA 91125, USA}
\email{matilde\@@caltech.edu}
\email{teh\@@caltech.edu}
\address{Department of Physics and Astronomy \\
University of Southern California \\
Los Angeles, CA 90089, USA}
\email{pierpaol\@@usc.edu}

\begin{document}
\maketitle

\begin{abstract}
We show that, in a model of modified gravity based on the spectral action functional,
there is a nontrivial coupling between cosmic topology and inflation, in the sense that
the shape of the possible slow-roll inflation potentials obtained in the model from the
nonperturbative form of the spectral action are sensitive not only to the geometry (flat or 
positively curved) of the universe, but also to the different possible non-simply connected 
topologies. We show this by explicitly computing the nonperturbative spectral action for 
some candidate flat cosmic topologies given by Bieberbach manifolds and showing  
that the resulting inflation potential differs from that of the flat torus by a multiplicative
factor, similarly to what happens in the case of the spectral action of the spherical
forms in relation to the case of the $3$-sphere. We then show that, while the
slow-roll parameters differ between the spherical and flat manifolds but do not
distinguish different topologies within each class, the power spectra detect the
different scalings of the slow-roll potential and therefore distinguish between the
various topologies, both in the spherical and in the flat case.
\end{abstract}

\section{Introduction}

Noncommutative cosmology is a new and rapidly developing area of research, which
aims at building cosmological models based on a ``modified gravity" action functional
which arises naturally in the context of noncommutative geometry, the {\em spectral
action} functional of \cite{CC}. As we discuss more in detail in \S \ref{NCcosmSec}
below, this functional recovers the usual Einstein--Hilbert action, with additional terms,
such as a conformal gravity, Weyl curvature term. It also has the advantage of
allowing for interesting couplings of gravity to matter, when extended from manifolds
to ``almost commutative geometries"  as in \cite{CoSM} and later models \cite{CCM},
\cite{BroSu}. Thus, this approach makes it possible to recover from the same spectral action functional, in
addition to the gravitational terms, the full Lagrangian of various particle physics models, 
ranging from the Minimal Standard Model of \cite{CoSM}, to the extension with right
handed neutrinos and Majorana mass terms of \cite{CCM}, and to supersymmetric
QCD as in \cite{BroSu}. 
The study of cosmological models derived from the spectral action gave rise to
early universe models as in \cite{MaPie} and \cite{KoMa}, which present
various possible inflation scenarios, as well as effects on primordial black holes 
evaporation and gravitational wave propagation. Effects on gravitational waves,
as well as inflation scenarios coming from the spectral action functional, were also
recently studied in \cite{NeOSa1}, \cite{NeOSa2}, \cite{NeSa}.

Our previous work \cite{MaPieTeh} showed that, when one considers the nonperturbative
form of the spectral action, as in \cite{CC2}, one obtains a slow-roll potential for inflation.
We compared some of the more likely candidates for cosmic topologies (the quaternionic
and dodecahedral cosmology, and the flat tori) and we showed that, in the spherical cases
(quaternionic and dodecahedral), the nonperturbative spectral action is just a multiple of
the spectral action of the sphere $S^3$, and consequently the inflation potential only
differs from the one of the sphere case by a constant scaling factor,  which cancels out
in the computation of the slow-roll parameters, which are therefore the same as in the 
case of a simply connected topology and do not distinguish the different cosmic topologies 
with the same spherical geometry.

This result for spherical space forms was further confirmed and extended in \cite{Teh}, where the nonperturbative spectral action is computed explicitly  for {\em all} the spherical space forms 
and it is shown to be always a multiple of the spectral action of $S^3$, with a proportionality
factor that depends explicitly on the 3-manifold.
Thus, different candidate cosmic topologies with the same positively curved geometry 
yield the same values of the slow-roll parameters and of the power-law indices and 
tensor-to-scalar ratio, which are computed from these parameters.

In \cite{MaPieTeh}, however, we showed that the inflation potential obtained from the nonperturbative spectral action is different in the case of the flat tori, and not just by
a scalar dilation factor. Thus, we know already that the possible inflation scenarios in 
noncommutative cosmology depend on the underlying geometry (flat
or positively curved) of the universe, and the slow--roll parameters are different for
these two classes.

The slow-roll parameters alone only
distinguish, in our model, between the flat and spherical geometries but
not between different topologies within each class. However,
in the present paper we show that, when one looks at the
amplitudes for the power spectra for density perturbations and gravitational
waves (scalar and tensor perturbations), these detect the different scaling
factors in the slow-roll potentials we obtain for the different spherical and
flat topologies, hence we obtain genuinely different inflation scenarios 
for different cosmic topologies. 

We achieve this result by relying on the computations of the nonperturbative spectral action,
which in the spherical cases are obtained in \cite{MaPieTeh} and \cite{Teh},  and by
deriving in this paper the analogous explicit computation 
of the nonperturbative spectral action for the flat Bieberbach manifolds.
A similar computation of the spectral action for Bieberbach manifolds
was simultaneously independently obtained by Piotr Olczykowski and 
Andrzej Sitarz in \cite{OlSi}.

Thus, the main conclusion of this paper is that {\em a modified gravity
model based on the spectral action functional predicts a coupling between 
cosmic topology and inflation potential, with different scalings in the
power spectra that distinguish between different topologies, and 
slow-roll parameters that distinguish between the spherical and
flat cases.}

The paper is organized as follows.
We first describe in \S \ref{InflGeomSec} the broader context in which the problem
we consider here falls, namely the cosmological results
relating inflation, the geometry of the universe, and the background radiation,
and the problem of cosmic topology.
We then review briefly in \S \ref{NCcosmSec} the use of the spectral action as
a modified gravity functional and the important distinction between its asymptotic
expansion at large energies and the nonperturbative form given in terms of
Dirac spectra. 
In \S \ref{BiebSec} we present the main mathematical result of this
paper, which gives an explicit calculation of
the nonperturbative spectral action for certain Bieberbach manifolds, using the
Dirac spectra of \cite{Pfa} and a Poisson summation technique similar to that
introduced in \cite{CC2},  and used in \cite{MaPieTeh} and \cite{Teh}. 

Finally, in \S \ref{InflPotSec} we compare the resulting slow-roll inflation potentials,
power spectra for density perturbations and slow-roll parameters, for all the 
different possible cosmic topologies.

\subsection{Inflation, geometry, and topology}\label{InflGeomSec}

It is well known that the mechanism of cosmic inflation, first proposed by Alan Guth and
Andrei Linde, naturally leads to a flat or almost flat geometry of the universe (see for
instance \S 1.7 of \cite{Linde}). It was then shown in \cite{KaSpeSu} that the geometry
of the universe can be read in the cosmic microwave background radiation (CMB), by
showing that the anisotropies of the CMB depend primarily upon the geometry 
of the universe (flat, positively or negatively curved) and that this information can
be detected through the fact that the location of the first Doppler peak changes for
different values of the curvature and is largely unaffected by other parameters. This
theoretical result made it possible to devise an observational test that could confirm
the inflationary theory and its prediction for a flat or nearly flat geometry. The experimental confirmation of the nearly flat geometry of the universe came in \cite{dBL} through the
Boomerang experiment. Thus, the geometry of the universe leaves
a measurable trace in the CMB, and measurements confirmed the flat geometry 
predicted by inflationary models.

The cosmic topology problem instead concentrates not on the question about
the curvature and the geometry of the universe, but on the possible existence,
for a given geometry, of a non-simply connected topology, that is, of whether
the spatial sections of spacetime can be compact 3-manifolds which are either
quotients of the 3-sphere (spherical space forms) in the positively curved case,
quotients of 3-dimensional Euclidean space (flat tori or Bieberbach manifolds)
in the flat case, or quotients of the 3-dimensional hyperbolic space (hyperbolic
3-manifolds) in the negatively curved space. A general introduction to the
problem of cosmic topology is given in \cite{LaLu}.

Since the cosmological observations prefer a flat or nearly flat positively
curved geometry to a nearly flat negatively curved geometry (see \cite{dBL}, \cite{WDP}),
most of the work in trying to identify the most likely candidates for a non-trivial
cosmic topology concentrate on the flat spaces and the spherical space forms.
Various methods have been devised to try to detect signatures of cosmic topology
in the CMB, in particular through a detailed analysis of simulated CMB skies for various 
candidate cosmic topologies (see \cite{RWULL} for the flat cases).
It is believed that perhaps some puzzling features of the CMB
such as the very low quadrupole, the very planar octupole, and the 
quadrupole--octupole alignment may find an explanation in the possible
presence of a non-simply connected topology, but no conclusive results to
that effect have yet been obtained. 

The recent results of \cite{MaPieTeh} 
show that a modified gravity model based on the
spectral action functional imposes constraints on the form of the possible
inflation slow-roll potentials, which depend on the geometry and topology
of the universe, as shown in \cite{MaPieTeh}. While the
resulting slow-roll parameters and spectral index and tensor-to-scalar
ratio  distinguish the even very slightly positively curved case from the
flat case, these parameters alone do not distinguish between the 
different spherical topologies, as shown in \cite{Teh}. As we show in
this paper, the situation is similar for the flat manifolds: these same
parameters alone do not distinguish between the various Bieberbach
manifolds (quotients of the flat torus), but they do distinguish these
from the spherical quotients. However, if one considers, in addition
to the slow-roll parameters, also the power spectra for the density
fluctuations, one can see that, in our model based on the 
spectral action as a modified gravity functional, the resulting
slow-roll potentials give different power spectra that distinguish between 
all the different topologies. 

\subsection{Slow-roll potential and power spectra of fluctuations}

We first need to recall here some well known facts about slow-roll inflation
potentials, slow-roll parameters, and the power spectra for density perturbations
and gravitational waves.
We refer the reader to \cite{SKamCoo} and to \cite{Lidsey}, \cite{StLy}, as well
as to the survey of inflationary cosmology \cite{Baumann}.

Consider an expanding universe, which is topologically a cylinder $Y\times \R$,
for a $3$-manifold $Y$, with a Lorentzian metric of the usual Friedmann form
\begin{equation}\label{Friedmann}
ds^2 = - dt^2 + a(t)^2 ds_Y^2  
\end{equation}
where $ds_Y^2= g_{ij} dx^i dx^j$ is the Riemannian metric on the $3$-manifold $Y$.

In models of inflation based on a single scalar field slow-roll potential $V(\phi)$, the dynamics
of the scale factor $a(t)$ in the Friedmann metric \eqref{Friedmann} is related to the
scalar field dynamics through the acceleration equation 
\begin{equation}\label{accel}
\frac{\ddot{a}}{a} = H^2 (1-\epsilon),
\end{equation}
where $H$ is the Hubble parameter, which is related to the scalar field and the
inflation potential by
\begin{equation}\label{Hubble}
H^2 = \frac{1}{3} \left( \frac{1}{2} \dot{\phi}^2 + V(\phi)\right), \ \ \text{ and }
\ \  \ddot{\phi} + 3 H \dot{\phi} + V^\prime(\phi) =0,
\end{equation}
and $\epsilon$ is the slow-roll parameter, which depends on the potential $V$
as described in \eqref{slowrollparam} below, see \cite{Baumann} for more details.
 
It is customary to decompose perturbations of the metric $ds^2$ of \eqref{Friedmann} 
into scalar and tensor perturbations, which correspond, respectively, to density 
fluctuations and gravitational waves. One typically neglects the remaining vector
components of the perturbation, assuming that these are not generated by
inflation and decay with the expansion of the universe, see \S 9.2 of \cite{Baumann}. 
Thus, one writes scalar and tensor perturbations in the form
\begin{equation}\label{scaltensper}
ds^2 =  - (1+2\Phi) dt^2 + 2 a(t) \, dB \, dt + a(t)^2 ((1-2\Psi) g_{ij}+2\Delta E + h_{ij}) dx^i dx^j
\end{equation}
with $dB=\partial_i B dx^i$ and $\Delta E= \partial_i\partial_j E$, 
and where the $h_{ij}$ give the tensor part of the perturbation, satisfying $\partial^i h_{ij}=0$ 
and $h^i_i=0$. The tensor perturbations $h_{ij}$ have two polarization modes, which
correspond to the two polarizations of the gravitational waves.

One considers then the  intrinsic curvature perturbation 
\begin{equation}\label{intrcurv}
\cR = \Psi - \frac{H}{\dot{\phi}}\delta \phi,
\end{equation}
which measures the spatial curvature of a comoving hypersurface, that is, a hypersurface with
constant $\phi$. After expanding $\cR$ in Fourier modes in the form
\begin{equation}\label{Rfourier}
\cR = \int \frac{d^3k}{(2\pi)^{3/2}} \cR_k\, e^{ikx} ,
\end{equation}
one obtains the power spectrum $\cP_s(k)$ for the density fluctuations (scalar perturbations
of the metric) from the two-point correlation function,
\begin{equation}\label{Ps2point}
 \langle \cR_k \cR_{k'} \rangle = (2\pi^2)^3\, \cP_s(k)\, \delta^3(k+k') .
\end{equation}
In the case of a Gaussian distribution, the power spectrum describes the complete
statistical information on the perturbations,  
while the higher order correlations functions contain the information on the possible
presence of non-Gaussianity phenomena. The power spectrum $\cP_t(k)$ for the
tensor perturbations is similarly obtained by expanding the tensor fluctiations in
Fourier modes $h_k$ and computing the two-point correlation function
\begin{equation}\label{tens2point}
\langle h_k h_{k'} \rangle = (2\pi^2)^3\, \cP_t(k)\, \delta^3(k+k').
\end{equation}
See \cite{Baumann}, \S 9.3, and \cite{StLy} for more details.

In slow-roll inflation models, the power spectra $\cP_s(k)$ and $\cP_t(k)$ are related
to the slow-roll potential $V(\phi)$ through the leading order expression (see \cite{SKamCoo})
\begin{equation}\label{PstV}
\cP_s(k) \sim \frac{1}{M_{Pl}^6} \frac{V^3}{(V^\prime)^2} \ \  \text{ and } \ \ 
\cP_t(k) \sim \frac{V}{M_{Pl}^4},
\end{equation}
up to a constant proportionality factor, and with $M_{Pl}$ the Planck mass.
Here the potential $V(\phi)$ and its derivative $V^\prime(\phi)$ are to be
evaluated at $k= a H$ where the corresponding scale leaves the horizon
during inflation. These can be expressed a power law as (\cite{SKamCoo})
\begin{equation}\label{powerlawP}
\begin{array}{rl}
\cP_s(k) \sim & \cP_s(k_0) 
\displaystyle{\left(\frac{k}{k_0} \right)^{1 - n_s + \frac{\alpha_s}{2} \log(k/k_0)}}
\\[3mm]  \cP_t(k) \sim & \cP_t(k_0) \displaystyle{\left(\frac{k}{k_0} 
\right)^{n_t + \frac{\alpha_t}{2} \log(k/k_0)}} , \end{array}
\end{equation}
where the spectral parameters $n_s$, $n_t$, $\alpha_s$, and $\alpha_t$
depend on the slow--roll potential in the following way.
In the slow-roll approximation, the slow-roll parameters are given by the expressions
\begin{equation}\label{slowrollparam}
\begin{array}{rl}
\epsilon = &  \displaystyle{\frac{M_{Pl}^2}{16\pi} \left( \frac{V^\prime}{V} \right)^2} \\[3mm]
\eta = & \displaystyle{\frac{M_{Pl}^2}{8\pi} \frac{V^{\prime\prime}}{V}} \\[3mm]
\xi = & \displaystyle{\frac{M_{Pl}^4}{64\pi^2} \frac{V^\prime V^{\prime\prime\prime}}{V^2}} 
\end{array}
\end{equation}
Notice that we follow here a different convention with respect to the one
we used in \cite{MaPieTeh} on the form of the slow-roll parameters.
The spectral parameters are then obtained from these as
\begin{equation}\label{spectralparam}
\begin{array}{rl}
n_s \simeq & 1 - 6 \epsilon + 2 \eta \\[2mm]
n_t \simeq & - 2 \epsilon \\[2mm]
\alpha_s \simeq & 16 \epsilon \eta - 24 \epsilon^2 - 2 \xi \\[2mm]
\alpha_t \simeq & 4 \epsilon\eta - 8 \epsilon^2
\end{array}
\end{equation}
while the tensor-to-scalar ratio is given by
\begin{equation}\label{tensorscalar}
r = \frac{P_t}{P_s} = 16 \epsilon.
\end{equation}

{}From the point of view of our model, the following observation will be useful
when we compare the slow-roll potentials that we obtain for different cosmic
topologies and how they affect the power spectra.

\begin{lem}\label{scaleVandP} Given a slow-roll potential $V(\phi)$
and the corresponding power spectra $\cP_s(k)$ and $\cP_t(k)$ as in
\eqref{PstV} and \eqref{powerlawP}. If the potential $V(\phi)$ is rescaled
by a constant factor $V(\phi) \mapsto \lambda V(\phi)$, then the power
spectra $\cP_s(k)$ and $\cP_t(k)$ are also rescaled by the same factor 
$\lambda >0$, while in the power law \eqref{powerlawP} the exponents
are unchanged.
\end{lem}

\proof This is an immediate consequence of \eqref{PstV},  \eqref{slowrollparam},
\eqref{spectralparam}, and \eqref{powerlawP}. In fact, from \eqref{PstV}, we
see that $V(\phi) \mapsto \lambda V(\phi)$ maps $\cP_t \mapsto \lambda \cP_t$,
and also $\cP_s \mapsto \lambda \cP_s$, since it transforms $V^3 (V^\prime)^{-2} \mapsto
\lambda V^3 (V^\prime)^{-2}$. On the other hand, the expressions $(V^\prime /V)^2$ and
$V^{\prime\prime}/V$ and $V^\prime V^{\prime\prime\prime}/V^2$ in the slow-roll parameters
\eqref{slowrollparam} are left unchanged by $V \mapsto \lambda V$, so that the 
slow-roll parameters and all the resulting spectral parameters of \eqref{spectralparam} 
are unchanged. Thus, the power law \eqref{powerlawP} only changes by a multiplicative
factor $\cP_s(k_0)\mapsto \lambda \cP_s(k_0)$ and $\cP_t(k_0)\mapsto \lambda \cP_t(k_0)$,
with unchanged exponents.
\endproof

\section{Noncommutative cosmology}\label{NCcosmSec}

\subsection{The spectral action as a modified gravity model}

In its nonperturbative form, the spectral action is defined in terms of
the spectrum of the Dirac operator, on a spin manifold or more generally
on a noncommutative space (a spectral triple), as the functional $\Tr( f(D/\Lambda) )$,
where $f$ is a smooth test function and $\Lambda$ is an energy scale that
makes $D/\Lambda$ dimensionless. 

The reason why this can be regarded as an action functional for gravity (or
gravity coupled to matter in the noncommutative case) lies in the fact that,
for large energies $\Lambda$ it has an asymptotic expansion (see \cite{CC})
of the form
\begin{equation}\label{SpActAsympt}
 \Tr(f(D/\Lambda))\sim \sum_{k\in {\rm DimSp^+}} f_{k} \Lambda^k {\int\!\!\!\!\!\!-} |D|^{-k} + f(0) \zeta_D(0)+ o(1),
\end{equation}
with $f_k= \int_0^\infty f(v) v^{k-1} dv$ and with the integrations
$$  {\int\!\!\!\!\!\!-} |D|^{-k} $$
given by residues of zeta function $\zeta_D (s) = \Tr (|D|^{-s})$ at
the positive points of the {\em dimension spectrum} of the spectral triple,
that is, the set of poles of the zeta functions. In the case of a 4-dimensional spin manifold,
these, in turn, are expressed in terms of integrals of curvature terms. These include
the usual Einstein--Hilbert action
$$ \frac{1}{2\kappa_0^2}  \int\,R
 \, \sqrt g \,d^4 x $$
and a cosmological term
$$  \gamma_0 \,\int \,\sqrt g\,d^4 x, $$
but it also contains 
some additional terms, like a non-dynamical
topological term 
$$  \tau_0 \int R^* R^* \sqrt g \,d^4 x, $$
where $R^*R^*$ denotes the form that represents the Pontrjagin class and
integrates to a multiple of the Euler characteristic of the manifold,
as well as a conformal gravity term 
$$  \alpha_0 \int C_{\mu
\nu \rho \sigma} \, C^{\mu \nu \rho \sigma} \sqrt g \,d^4 x, $$
which is given in terms of the Weyl curvature tensor. We do not
give any more details here and we refer the reader to Chapter 1 of
\cite{CoMa} for a more complete treatment. 

The presence of conformal gravity terms along with the
Einstein--Hilbert and cosmological terms give then a modified
gravity action functional. When one considers the nonperturbative
form of the spectral action, rather than its asymptotic expansion
at large energies, one can find additional nonperturbative
correction terms. One of these was identified in \cite{CC2}, in
the case of the 3-sphere, as a potential for a scalar field, which
was interpreted in \cite{MaPieTeh} as a potential for a cosmological
slow-roll inflation scenario, and computed for other, non-simply
connected cosmic topologies.

\section{Geometry, topology and inflation: spherical forms}\label{SpherePowerSec}

The nonperturbative spectral action for the spherical space forms $Y=S^3/\Gamma$
was computed recently by one of the authors \cite{Teh}. It turns out that, although
the Dirac spectra can be significantly different for different spin structures, the spectral
action itself is independent of the choice of the spin structure, and it is always equal
to a constant multiple of the spectral action for the 3-sphere $S^3$, where the
multiple is just dividing by the order of the group $\Gamma$. This is exactly what
one expects by looking at the asymptotic expansion of the spectral action for
large energies $\Lambda$, and the only significant nonperturbative effect arises
in the form of a slow-roll potential, as in \cite{CC2}, \cite{MaPieTeh}.

\begin{thm}\label{sphereformsSpAc} {\em (Teh, \cite{Teh})}  For all the
spherical space forms $Y = S^3/\Gamma$ with the round metric induced from $S^3$,
and for all choices of spin structure,  the nonperturbative spectral action on $Y$ is equal to
\begin{equation}\label{SpAcSphereForm}
\Tr(f(D_Y/\Lambda)) = \frac{1}{\# \Gamma} \left( \Lambda^3 \widehat{f}^{(2)}(0) - \frac{1}{4} \Lambda \widehat{f}(0) \right) = \frac{1}{\# \Gamma} \Tr(f(D_{S^3}/\Lambda)),
\end{equation}
up to order $O(\Lambda^{-\infty})$.
\end{thm}

Correspondingly, as explained in \S 5 of \cite{MaPieTeh}, one obtains a slow-roll
potential by considering the variation $D^2 \mapsto D^2 +\phi^2$
of the spectral action as in \cite{CC2}. More precisely, one considers a Euclidean
compactification of the 4-dimensional spacetime $Y \times \R$ to a compact
Riemannian manifold $Y \times S^1$ with the compactification $S^1$ of size
$\beta$. One then computes the spectral action $\Tr(h(D^2_{Y\times S^1}/\Lambda^2))$
on this compactification and its variation
\begin{equation}\label{Spact4Dphi}
\Tr(h((D^2_{Y\times S^1}+\phi^2)/\Lambda^2)) - \Tr(h(D^2_{Y\times S^1}/\Lambda^2)) = V_Y(\phi),
\end{equation}
up to terms of order $O(\Lambda^{-\infty})$, where the potential $V(\phi)$ is given 
by the following.

\begin{prop}\label{S3potential}
Let $Y$ be a spherical space form $Y=S^3/\Gamma$ with the induced round metric. Let $a>0$
be the radius of the sphere $S^3$ and $\beta$ the size of the circle $S^1$ in the Euclidean
compactification $Y\times S^1$. Then the slow-roll potential $V(\phi)$ in \eqref{Spact4Dphi}
is of the form
\begin{equation}\label{VphiVW}
V_Y(\phi) = \pi \Lambda^4 \beta a^3 \cV_Y (\frac{\phi^2}{\Lambda^2}) + \frac{\pi}{2} \Lambda^2 \beta a \cW_Y(\frac{\phi^2}{\Lambda^2}),
\end{equation}
where 
\begin{equation}\label{VWYS3}
\cV_Y(x) = \lambda_Y \, \cV_{S^3}(x) \ \  \text{ and } \ \  \cW_Y(x) = \lambda_Y \, \cW_{S^3}(x),
\end{equation}
with
\begin{equation}\label{lambdaY}
\lambda_Y  = \frac{1}{\# \Gamma} \ \ \ \text{ for } \ \  Y = S^3 /\Gamma,
\end{equation}
and with
\begin{equation}\label{VWS3}
\cV_{S^3}(x) = \int_0^\infty u \, (h(u+x) - h(u))\, du  \ \ \  \text{ and } \ \ \ 
\cW_{S^3}(x) = \int_0^x h(u)\, du.
\end{equation}
\end{prop}

\proof The statement follows directly from the result of Theorem 7 of \cite{CC2} and
\S 5 of \cite{MaPieTeh}. \endproof

In particular, for the different spherical forms, the potential has the same form as
that of the 3-sphere case, but it is scaled by the factor $\lambda_Y  = 1/\# \Gamma$,
\begin{equation}\label{VYS3}
V_Y(\phi) = \lambda_Y \, V_{S^3}(\phi) = \frac{V_{S^3}(\phi)}{\# \Gamma}.
\end{equation}

Notice, moreover, that in the potential $V_{S^3}(\phi)$ one has an overall factor
of $(\Lambda a)^3 (\Lambda \beta)$ that multiplies the $\cV_{S^3}$ term and 
a factor of $(\Lambda a)(\Lambda \beta)$ that multiplies the $\cW_{S^3}$ term.
As we observed already in \cite{MaPieTeh}, when one Wick rotates back to
the Minkowskian model with the Friedmann metric, both the scale factor $a(t)$
and the energy scale $\Lambda(t)$ evolve with the expansion of the universe,
but in such a way that $\Lambda(t) \sim 1/a(t)$ so that the product $\Lambda a\sim 1$. 
In \cite{MaPieTeh} we did not need to analyze the behavior of
the $\Lambda \beta$ factor, since we only looked at the slow-roll parameters
\eqref{slowrollparam} where that factor cancels out. In the spectral action model
of cosmology, the choice of the scale $\beta$ of the Euclidean compactification
is an artifact of the model, which allows one to compute the spectral action
in terms of the spectrum of the Dirac operator on the compact Riemannian
4-manifold $Y \times S^1$. Eventually, the physically significant quantities
derived from the spectral action functional are Wick rotated back to the
Minkowskian signature case. Since in its nonperturbative form the spectral
action functional is supposed to give a modified gravity action functional
that works at all scales, not just in the asymptotic expansion for large
$\Lambda$, it seems therefore natural to set the choice of the length $\beta$
in the model so that the product $\Lambda\beta \sim 1$ remains constant. 

Another reason for making the assumption that $\Lambda\beta \sim 1$ is the
interpretation given in \cite{CC2} of the parameter $\beta$ in the Euclidean
compactification as an inverse temperature. Then, up to a universal constant,
that behaves like the inverse of an energy scale and, when rotating back
to the Minkowskian signature, one knows that, in the expansion of the universe
the scale factor $a(t)$ is inversely proportional to the temperature, so that
the assumption $\Lambda\beta \sim 1$ is justified.

With this setting, the slow-roll potential one obtains in the case of the 3-sphere is
of the form
\begin{equation}\label{VS3nobeta}
V_{S^3}(\phi) = \pi \int_0^\infty u \, (h(u+x) - h(u))\, du + \frac{\pi}{2} \int_0^x h(u)\, du.
\end{equation}

Then one has the following result for the power spectra for the various cosmic topology
candidates given by spherical space forms.

\begin{prop}\label{PowerSpheres}
Let $\cP_{s,Y}(k)$ and $\cP_{t,Y}(k)$ denote the power spectra for the density
fluctuations and the gravitational waves, computed as in \eqref{PstV}, for the 
slow-roll potential $V_Y(\phi)$. Then they satisfy the power law
\begin{equation}\label{powerlawPY}
\begin{array}{rl}
\cP_{s,Y}(k) \sim & \lambda_Y \, \cP_s(k_0) 
\displaystyle{\left(\frac{k}{k_0} \right)^{1 - n_{s,S^3} + \frac{\alpha_{s,S^3}}{2} \log(k/k_0)}}
\\[3mm]  \cP_{t,Y}(k) \sim & \lambda_Y \, \cP_t(k_0) \displaystyle{\left(\frac{k}{k_0} 
\right)^{n_{t,S^3} + \frac{\alpha_{t,S^3}}{2} \log(k/k_0)}} , \end{array}
\end{equation}
where $\lambda_Y = 1/\#\Gamma$ for $Y=S^3/\Gamma$ and the spectral parameters
$n_{s,S^3}$, $n_{t,S^3}$, $\alpha_{s,S^3}$, $\alpha_{t,S^3}$ are computed as in
\eqref{spectralparam} from the slow-roll parameters \eqref{slowrollparam}, which satisfy
$\epsilon_Y = \epsilon_{S^3}$, $\eta_Y=\eta_{S^3}$, $\xi_Y=\xi_{S^3}$.
\end{prop}

\begin{figure}
\includegraphics[scale=0.8]{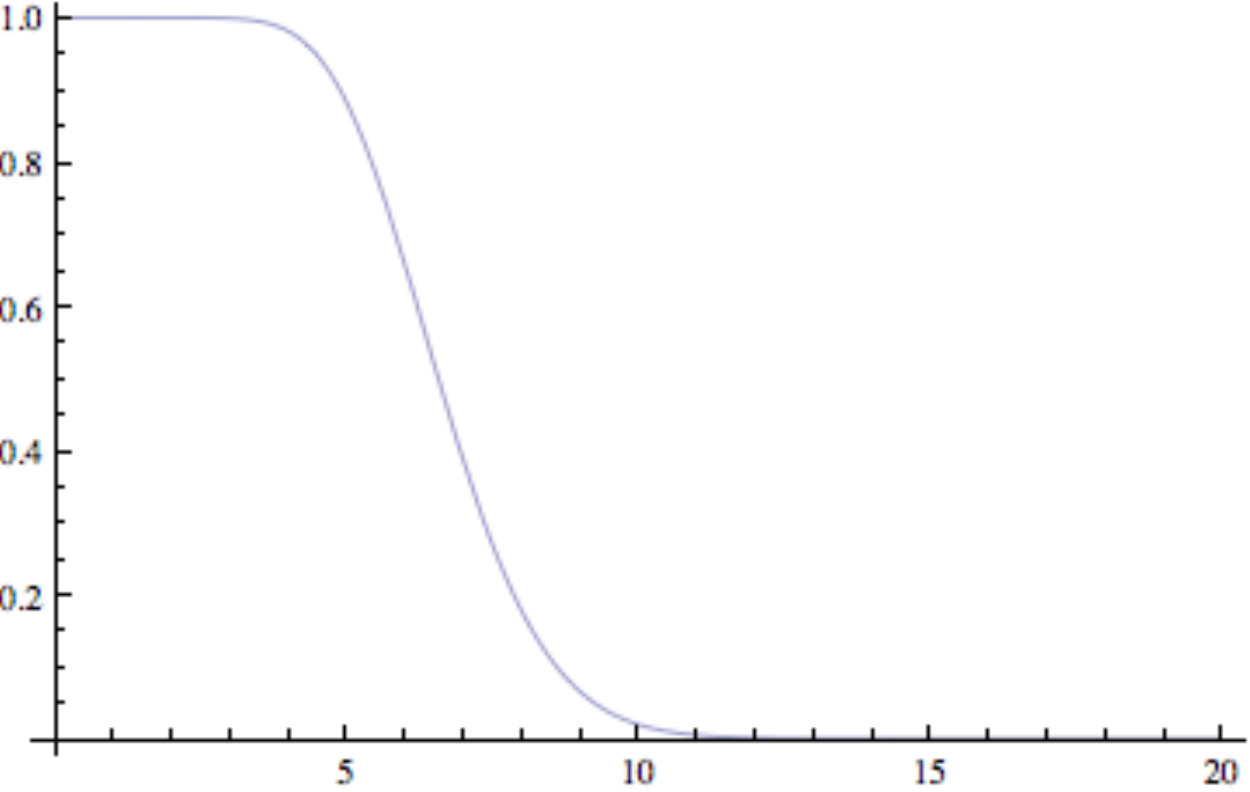}
\caption{The test function $h(x)=h_n(x)$ with $n=20$. \label{hnFig}}
\end{figure}

\begin{figure}
\includegraphics[scale=0.8]{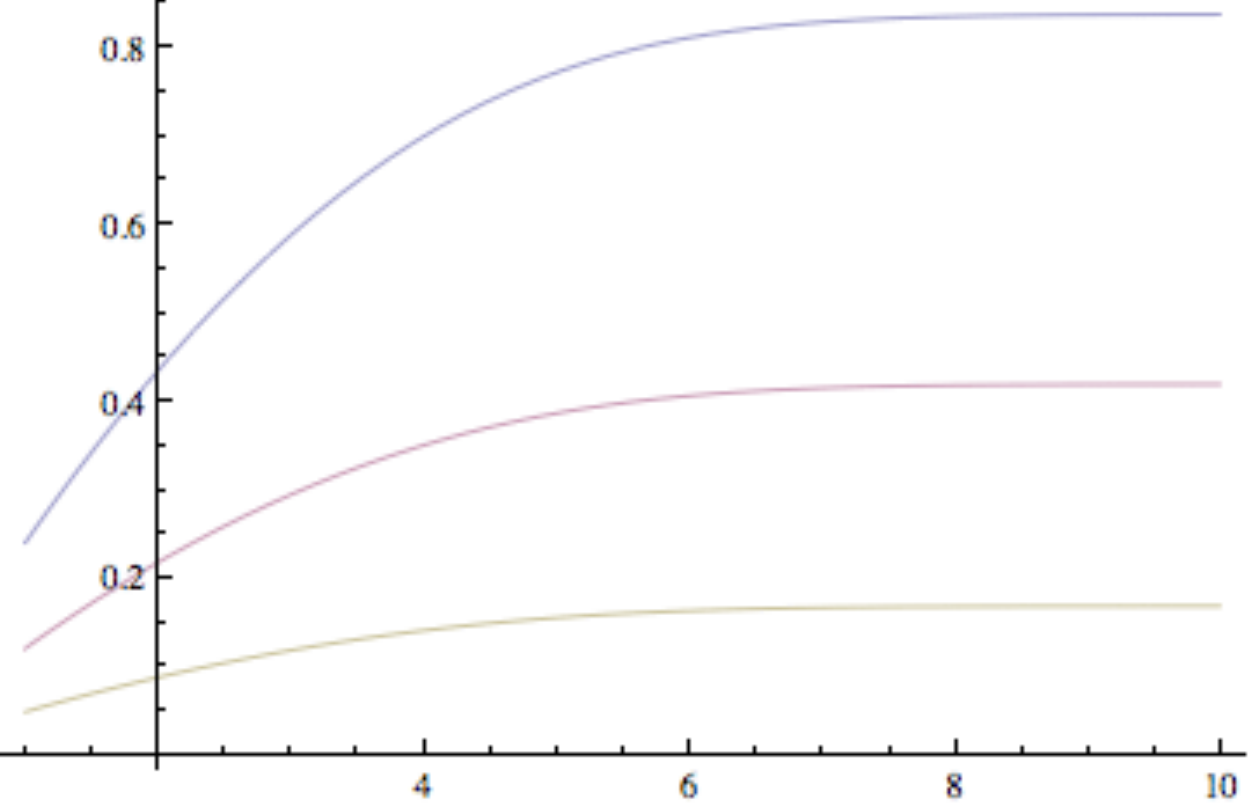}
\caption{The slow-roll potentials for the binary tetrahedral, binary octahedral, and binary icosahedral cases. \label{VsphericalFig}}
\end{figure}

To see explicitly the effect on the slow-roll potential of the scaling by
$\lambda_Y$, we consider the same test functions $h_n(x)$ used in
\cite{CC2} to approximate smoothly a cutoff function. These are given by
$$ h_n(x) = \sum_{k=0}^n \frac{(\pi x)^k}{k!} e^{-\pi x}. $$
Figure \ref{hnFig} shows the graph of $h_n(x)$ when $n=20$.
We use this test function to compute the slow-roll potential using the function
$\cV(x) + \frac{1}{2} \cW(x)$, after setting the factors $\Lambda a=1$ and 
$\Lambda \beta =1$, and up to an overall multiplicative factor of $\pi$.
We then see in Figure \ref{VsphericalFig} the different curves of the slow-roll
potential for the three cases where $Y=S^3/\Gamma$ with $\Gamma$ the
binary tetrahedral, binary octahedral, or binary icosahedral
group, respectively given by the top, middle, and bottom curve.

\section{The spectral action for Bieberbach manifolds}\label{BiebSec}

We now consider the case of candidate cosmic topologies that are flat 3-manifolds.
The simplest case is the flat torus $T^3$, which we have already discussed in \cite{MaPieTeh}.
There are then the Bieberbach manifolds, which are obtained as quotients of the torus
by a finite group action. In this section we give an explicit computation of the nonperturbative
spectral action for the Bieberbach manifolds (with the exception of $G5$ which requires
a different technique and will be analyzed elsewhere), and in the next section we then
derive the analog of Proposition \ref{PowerSpheres} for the case of these flat geometries.

Calculations of the spectral action for Bieberbach manifolds were simultaneously
independently obtained in \cite{OlSi}.

\smallskip

The Dirac spectrum of Bieberbach manifolds is computed in \cite{Pfa} for each of the six affine equivalence classes of three-dimensional orientable Bieberbach manifolds, and for each possible choice of spin structure and choice of flat metric.  These classes are labeled $G1$ through $G6$, with $G1$ simply being the flat 3-torus.  

In general, the Dirac spectrum for each space depends on the choice of spin structure.  However, as in the case of the spherical manifolds, we show here that the nonperturbative spectral action
is {\em independent of the spin structure}.

We follow the notation of \cite{Pfa}, according to which the different possibilities for the Dirac spectra are indicated by a letter (e.g. $G2(a)$).  Note that it is possible for several spin 
structures to yield the same Dirac spectrum.

The nonperturbative spectral action for $G1$ was computed in \cite{MaPieTeh}.  We recall here the result for that case and then we restrict our discussion to the spaces $G2$ through $G6$.

\subsection{The structure of Dirac spectra of Bieberbach manifold}\label{BiebSpSec}

The spectrum of the Bieberbach manifolds generally consists of a symmetric component and an asymmetric component as computed in \cite{Pfa}.  The symmetric components are parametrized by subsets $I \subset \Z ^3$, such that the eigenvalues are given by some formula $\lambda_{x}$, $x \in I$, and the multiplicity of each eigenvalue, $\lambda$, is some constant times the number of $x \in I$ such that $\lambda = \lambda_x$.  In the case of $G2$, $G4$, $G5$, $G6$ the constant is $1$, while in the $G3$ case the constant is $2$.  

The approach we use here to compute the spectral action nonperturbatively consists of using the symmetries of $\lambda_x$ as a function of $x \in I$ to almost cover all of the points in $\Z^3$ and then apply the Poisson summation formula as used in \cite{CC2}.  By ``almost cover'', it is meant that it is perfectly acceptable if two-, one-, or zero-dimensional lattices through the origin are covered multiple times, or not at all.  

The asymmetric component of the spectrum appears only some of the time.  The appearance of the asymmetric component depends on the choice of spin structure.  For those cases where it appears, the eigenvalues in the asymmetric component consist of the set
\[
\cB = \left\{ 2 \pi \frac{1}{H}\left(k \mu + c \right) | \mu \in \Z \right\},
\]
where $c$ is a constant depending on the spin structure, and $k$ is given in the 
following table:

\begin{center}
\begin{tabular}{|c|c|} 
\hline
 Bieberbach manifold & $k$ \\
 \hline
$G2$ & $2$ \\ 
$G3$ & $3$ \\
$G4$ & $4$ \\
$G5$ & $6$ \\
\hline
\end{tabular}
\end{center}

For no choice of spin structure does $G6$ have an asymmetric component to its spectrum.
Each of the eigenvalues in $\cB$ has multiplicity $2$.  Using the Poisson summation formula as in \cite{CC2}, we see that the asymmetric component of the spectrum contributes to the spectral action
\begin{equation}
\label{asym}
\frac{\Lambda H}{\pi k}\int_{\R}f(u^2) du.
\end{equation}

The approach described here is effective for computing the nonperturbative spectral action for the manifolds labeled in \cite{Pfa} as $G2, G3, G4, G6$, but not for $G5$.  Therefore, we do not consider the $G5$ case in this paper: it will be discussed elsewhere.

\subsection{Recalling the torus case}\label{torusSec}

We gave in Theorem 8.1 of \cite{MaPieTeh} the explicit computation of the
non-perturbative spectral action for the torus. We recall here the statement
for later use.

\begin{thm}\label{SpAT3}
Let  $T^3 = \R^3/\Z^3$ be the flat torus with an arbitrary choice of spin structure.
The nonperturbative spectral action is of the form
\begin{equation}\label{T3spact}
\Tr (f(D^2/\Lambda^2)) = \frac{\Lambda^3}{4 \pi^3}  \int_{\R ^3} f(u^2 + v^2 + w^2)dudvdw ,
\end{equation}
up to terms of order $O(\Lambda^{-\infty})$.
\end{thm}

\subsection{The spectral action for $G2$}\label{G2sec}

The Bieberbach manifold $G2$ is the one that is described as ``half-turn space" in
the cosmic topology setting in \cite{RWULL}, because the identifications of the faces 
of the fundamental domain is achieved by introducing a $\pi$-rotation about the
$z$-axis.  It is obtained by considering a  lattice with basis $a_1=(0,0,H)$, 
$a_2=(L,0,0)$, and $a_3=(T,S,0)$, with $H,L,S \in \R^*_+$ and $T\in \R$,
and then taking the quotient $Y=\R^3/G2$ of $\R^3$ by the
group $G2$ generated by the commuting translations $t_i$ along these basis vectors 
$a_i$ and an additional generator $\alpha$ with relations 
\begin{equation}\label{G2rels}
\alpha^2 =t_1, \ \ \  \alpha t_2 \alpha^{-1} =t_2^{-1}, \ \ \  \alpha t_3 \alpha^{-1} = t_3^{-1} .
\end{equation}

Like the torus $T^3$, the Bieberbach manifold $G2$ has eight different spin structures,
parameterized by three signs $\delta_i=\pm 1$, 
see Theorem 3.3 of \cite{Pfa}.  Correspondingly, as shown in Theorem 5.7 of \cite{Pfa},
there are four different Dirac spectra, denoted $(a)$, $(b)$, $(c)$, and $(d)$, respectively
associated to the the spin structures 
\begin{center}
\begin{tabular}{|c|r|r|r|}
\hline
& $\delta_1$ & $\delta_2$ & $\delta_3$ \\ \hline
$(a)$ & $\pm 1$ & $1$ & $1$ \\ \hline
$(b)$ & $\pm 1$ & $-1$ & $1$ \\ \hline
$(c)$ & $\pm 1$ & $1$ & $-1$ \\ \hline
$(d)$ & $\pm 1$ & $-1$ & $-1$ \\ \hline
\end{tabular}
\end{center}

We give the computation of the nonperturbative spectral action separately
for each different spectrum and we will see that the result is independent 
of the spin structure and always a multiple of the spectral action of the torus.

\subsubsection{The case of $G2(a)$}\label{G2aSec}

In this first case, we go through the computation in full detail. 
The symmetric component of the spectrum is given by the data (\cite{Pfa})
$$I = \{(k,l,m)| k,l,m \in \Z, m \geq 1 \} \cup \{ (k,l,m) | k,l\in \Z, l\geq 1, m =0\}$$
$$\lambda_{klm}^{\pm} = \pm 2 \pi \sqrt{\frac{1}{H^2}(k+\frac{1}{2})^2+ \frac{1}{L^2}l^2 + \frac{1}{S^2}(m - \frac{T}{L}l)^2 },$$
We make the assumption that $T = L$.  Set $p = m-l$. Then we have equivalently:
$$I = \{(k,l,p)| k,l,p \in \Z, p > -l \} \cup \{ (k,l,p) | k,l\in \Z, l\geq 1, p = -l\} =: I_1 \cup I_2$$
$$\lambda_{klp}^{\pm} = \pm 2 \pi \sqrt{\frac{1}{H^2}(k+\frac{1}{2})^2+ \frac{1}{L^2}l^2 + \frac{1}{S^2}p^2 }.$$

\begin{thm}\label{thmG2a}
Let $G2(a)$ be the Bieberbach manifold $\R^3/G2$, with $T=L$ and with a
spin structure with $\delta_i=\{ \pm 1, 1,1\}$.
The nonperturbative spectral action of the manifold $G2(a)$ is of the form
\begin{equation}\label{G2aSA}
\Tr (f(D^2/\Lambda^2)) = H S L \left( \frac{\Lambda}{2 \pi} \right) ^3 \int_{\R ^3} f(u^2 + v^2 + w^2)dudvdw ,
\end{equation}
up to terms of order $O(\Lambda^{-\infty})$.
\end{thm}

\begin{figure}
\includegraphics[scale=0.4]{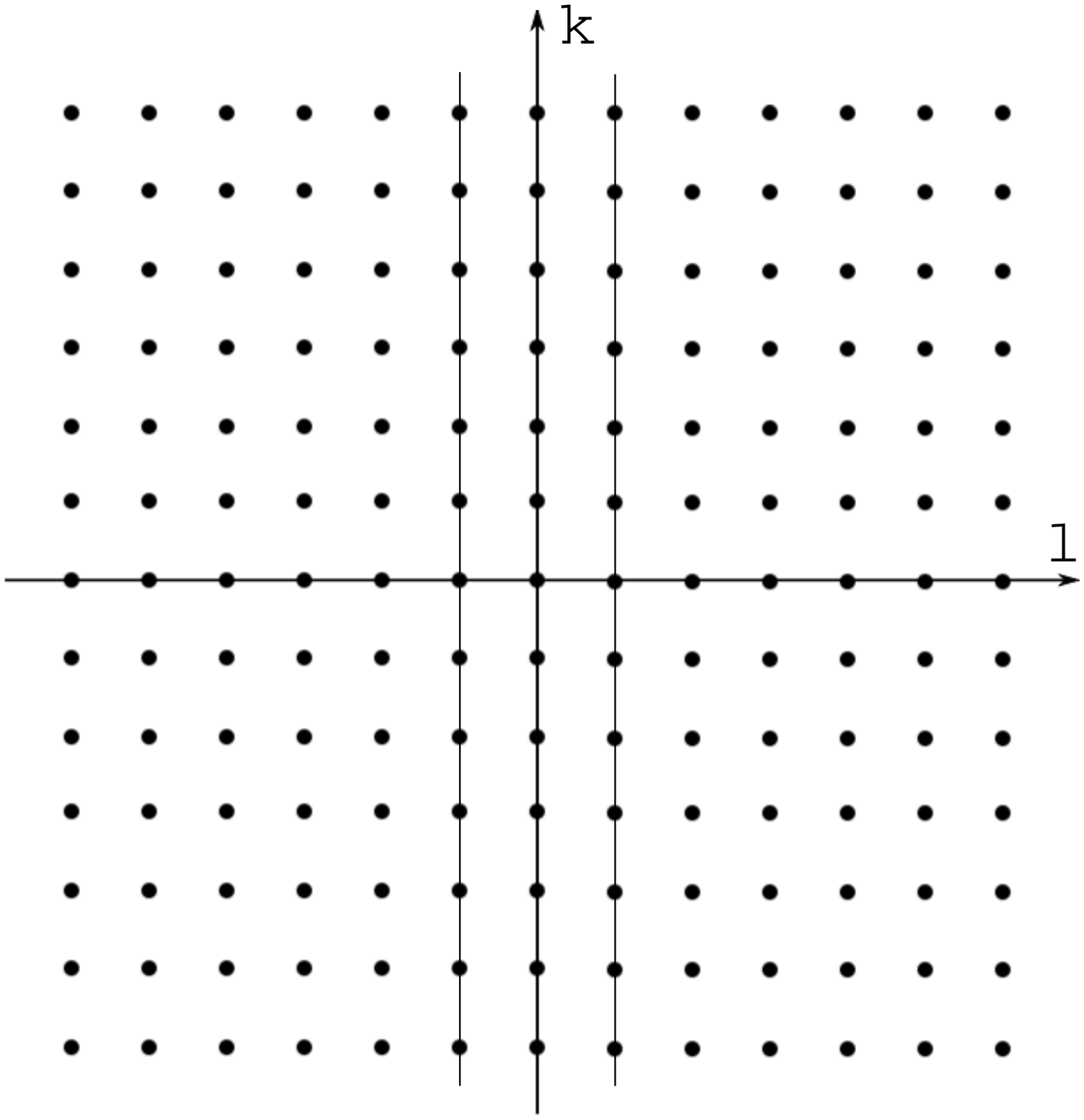}
\caption{Lattice decomposition for the $I_2$ contribution to the spectral action of $G2(a)$ Two regions and the set $l=0$.   \label{g2alkFig}}
\end{figure}

\proof We compute the contribution to the spectral action due to $I_1$. 
Since $\lambda_{klp} ^{\pm}$ is invariant under the transformation $l \mapsto -l$ and $p\mapsto -p$, we see that
\[
\sum_{\Z ^3}f(\lambda_{klp} ^2 / \Lambda^2) = 2 \sum_{I_1}f(\lambda_{klp} ^2 / \Lambda^2) + \sum_{p= -l}f(\lambda_{klp} ^2 / \Lambda^2).
\]
The decomposition of $\Z ^3$ used to compute this contribution to the spectral action is displayed in figure $\ref{g2alkFig}$.
Applying the Poisson summation formula we get a contribution to the spectral action of
\[
H S L \left( \frac{\Lambda}{2 \pi} \right) ^3 \int_{\R ^3} f(u^2 + v^2 + w^2) -  H \frac{LS}{\sqrt{L^2 + S^2}} \left( \frac{\Lambda}{2 \pi} \right)^2 \int_{R^2} f(u^2 + v^2) ,
\]
plus possible terms of order $O(\Lambda^{-\infty})$.

\begin{figure}
\includegraphics[scale=0.4]{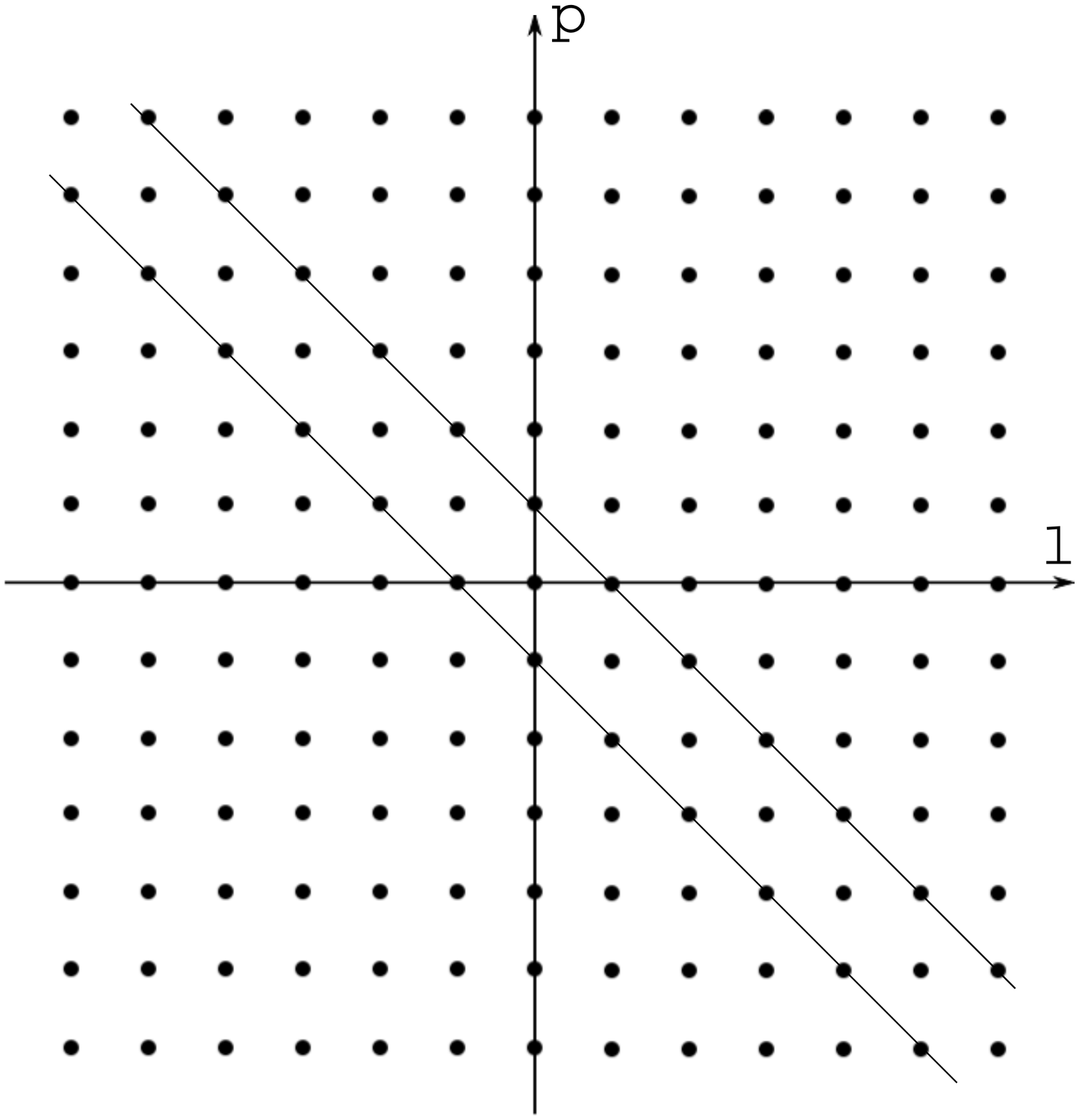}
\caption{Lattice decomposition for the $I_1$ contribution to the spectral action of $G2(a)$.  Two regions and the set $l=-p$  \label{g2alpFig}}
\end{figure}

As for $I_2$ we again use the fact that the spectrum is invariant under the transformation $l \mapsto -l$, $p \mapsto -p$ to see that
\[
\sum_{\Z ^2}f(\lambda_{kl(-l)} ^2 / \Lambda^2) = 2 \sum_{I_2}f(\lambda_{klp} ^2 / \Lambda^2) + \sum_{p= l=0}f(\lambda_{klp} ^2 / \Lambda^2).
\]
The decomposition for this contribution to the spectral action is displayed in figure $\ref{g2alpFig}$.
We get a contribution to the spectral action of
\[
H \frac{LS}{\sqrt{L^2 + S^2}} \left( \frac{\Lambda}{2 \pi} \right) ^2 \int_{\R ^2} f(u^2 + v^2) -  H \left( \frac{\Lambda}{2 \pi} \right) \int_{\R} f(u^2) 
\]
plus possible terms of order $O(\Lambda^{-\infty})$.

When we include the contribution \eqref{asym} due to the asymmetric component we see that the spectral action of the space G2-(a) is equal to
\[
\Tr f(D^2/\Lambda^2) = H S L \left( \frac{\Lambda}{2 \pi} \right) ^3 \int_{\R ^3} f(u^2 + v^2 + w^2)dudvdw 
\]
again up to possible terms of order $O(\Lambda^{-\infty})$.
\endproof

\subsubsection{The case of $G2(b)$ and $G2(d)$}\label{G2bdSec}

The spectra of $G2(b)$ and $G2(d)$ have no asymmetric component.  
The symmetric component is given by 
$$I = \{(k,l,m)| k,l,m \in \Z, l \geq 0 \}$$
$$\lambda_{klm}^{\pm} = \pm 2 \pi \sqrt{\frac{1}{H^2}(k+\frac{1}{2})^2+ \frac{1}{L^2}(l+\frac{1}{2})^2 + \frac{1}{S^2}(m + c - \frac{T}{L}(l+ \frac{1}{2}))^2 }.$$
Let us once again assume that $T=L$. 

\begin{thm}\label{thmG2bd}
Let $G2(b)$ and $G2(d)$ be the Bieberbach manifolds $\R^3/G2$, with $T=L$ and with a
spin structure with $\delta_i=\{ \pm 1, -1,1\}$ and $\delta_i=\{ \pm 1, -1,-1\}$, respectively.
The nonperturbative spectral action of the manifolds $G2(b)$ and $G2(d)$ is again of the form
\begin{equation}\label{G2bdSA}
\Tr (f(D^2/\Lambda^2) )= H S L \left( \frac{\Lambda}{2 \pi} \right) ^3 \int_{\R ^3} f(u^2 + v^2 + w^2)dudvdw ,
\end{equation}
up to terms of order $O(\Lambda^{-\infty})$.
\end{thm}

\begin{figure}
\includegraphics[scale=0.4]{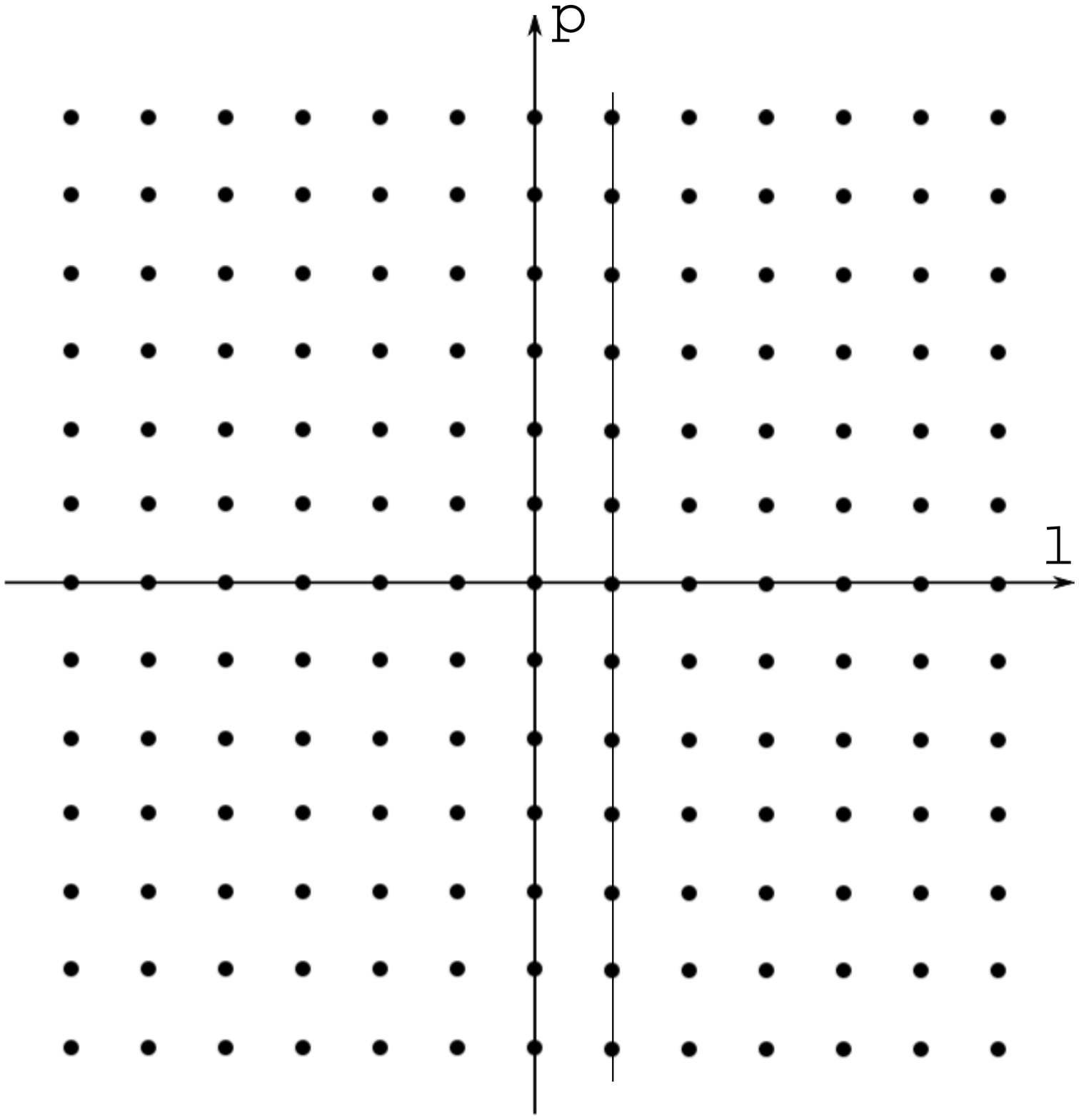}
\caption{Lattice decomposition for $G2(b),(d)$ computation.  Two regions.  \label{g2bdFig}}
\end{figure}

\proof With the assumption that $T=L$ and letting $p = m -l$, we can describe the spectrum equivalently by
$$I = \{(k,l,p)| k,l,p \in \Z, l \geq 0 \}$$
$$\lambda_{klp}^{\pm} = \pm 2 \pi \sqrt{\frac{1}{H^2}(k+\frac{1}{2})^2+ \frac{1}{L^2}(l+\frac{1}{2})^2 + \frac{1}{S^2}(p+ c + \frac{1}{2})^2 }.$$
Using the symmetry
\[
l \mapsto -1-l,
\]
we cover $\Z^3$ exactly, (see figure \ref{g2bdFig}) and we obtain the spectral action
\[
\Tr (f(D^2/\Lambda^2)) = H S L \left( \frac{\Lambda}{2 \pi} \right) ^3 \int_{\R ^3} f(u^2 + v^2 + w^2)dudvdw + O(\Lambda^{-\infty}).
\]
\endproof

\subsubsection{The case of $G2(c)$}\label{G2cSec}

In this case, the symmetric component of the spectrum is given by
\[
I = \{(k,l,m)| k,l,m \in \Z, m \geq 0 \}
\]
\[
\lambda_{klm}^{\pm} = \pm 2 \pi \sqrt{\frac{1}{H^2}(k+\frac{1}{2})^2+ \frac{1}{L^2}l^2 + \frac{1}{S^2}((m + 1/2) - \frac{T}{L}l)^2 }.
\]
Again, we assume $T = L$.  

\begin{thm}\label{thmG2c}
Let $G2(c)$ be the Bieberbach manifolds $\R^3/G2$, with $T=L$ and with a
spin structure with $\delta_i=\{ \pm 1, 1,-1\}$.
The nonperturbative spectral action of the manifold $G2(c)$ is again of the form
\begin{equation}\label{G2cSA}
\Tr (f(D^2/\Lambda^2) )= H S L \left( \frac{\Lambda}{2 \pi} \right) ^3 \int_{\R ^3} f(u^2 + v^2 + w^2)dudvdw ,
\end{equation}
up to terms of order $O(\Lambda^{-\infty})$.
\end{thm}

\begin{figure}
\includegraphics[scale=0.4]{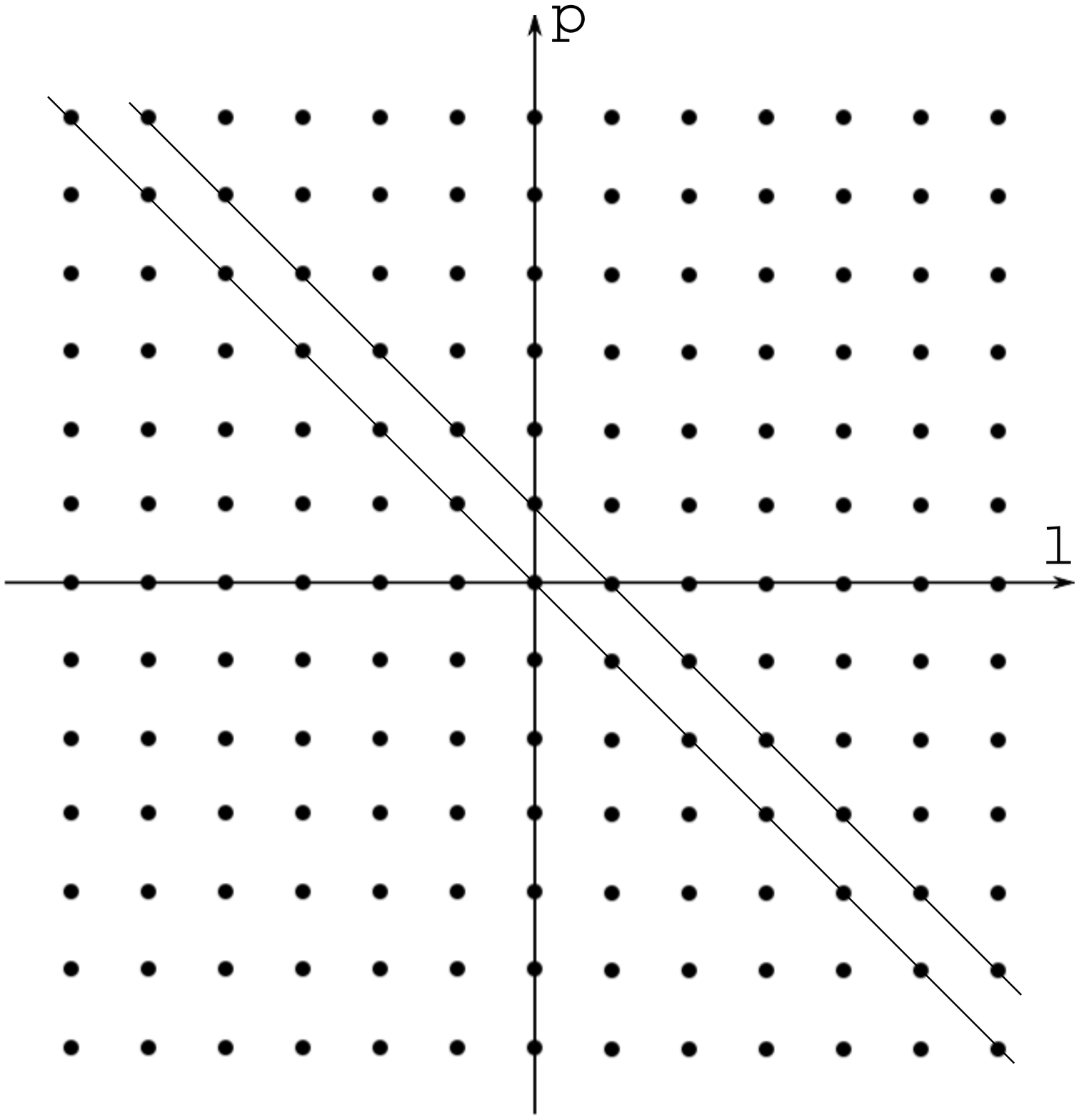}
\caption{Lattice decomposition for $G2(c)$ computation. Two regions.   \label{g2cFig}}
\end{figure}

\proof If we substitute $p = m-l$, we see that we may equivalently express the symmetric component with
\[
I = \{(k,l,p)| k,l,p \in \Z, p \geq  -l \}
\]
\[
\lambda_{klp}^{\pm} = \pm 2 \pi \sqrt{\frac{1}{H^2}(k+\frac{1}{2})^2+ \frac{1}{L^2}l^2 + \frac{1}{S^2}((p + 1/2)^2 }.
\]
Using the symmetry
\[
l \mapsto -l \quad p \mapsto 1-p,
\]
we cover $\Z^3$ exactly (see figure \ref{g2cFig}), and so the spectral action is again given by
\[
\Tr f(D^2/\Lambda^2) = H S L \left( \frac{\Lambda}{2 \pi} \right) ^3 \int_{\R ^3} f(u^2 + v^2 + w^2)dudvdw + O(\Lambda^{-\infty}).
\]
\endproof

\subsection{The spectral action for $G3$}\label{G3sec}

The Bieberbach manifold $G3$ is the one that, in the cosmic topology
setting of \cite{RWULL} is described as the ``third-turn space". One considers
the hexagonal lattice generated by vectors $a_1=(0,0,H)$, $a_2=(L,0,0)$
and $a_3=(-\frac{1}{2}L, \frac{\sqrt{3}}{2}L, 0)$, for $H$ and $L$ in $\R^*_+$,
and one then takes the quotient of $\R^3$ by the group $G3$ generated by
commuting translations $t_i$ along the vectors $a_i$ and an additional 
generator $\alpha$ with relations
\begin{equation}\label{relG3}
\alpha^3 = t_1,  \ \ \ \alpha t_2 \alpha^{-1} = t_3,  \ \ \  \alpha t_3 \alpha^{-1} = t_2^{-1} t_3^{-1}.  
\end{equation}
This has the effect of producing an identification of the faces of the fundamental
domain with a turn by an angle of $2\pi/3$ about the $z$-axis, hence the ``third-turn
space" terminology.

As shown in Theorem 3.3 of \cite{Pfa}, the Bieberbach manifold $G3$ has two
different spin structures, parameterized by one sign $\delta_1 =\pm 1$.  It is
then shown in Theorem 5.7 of \cite{Pfa} that these two spin structures have
different Dirac spectra, which are denoted as $G3(a)$ and $G3(b)$. We compute
below the nonperturbative spectral action in both cases and we show that, despite
the spectra being different, they give the same result for the nonperturbative
spectral action, which is again a multiple of the action for the torus.

\subsubsection{The case of $G3(a)$ and $G3(b)$}\label{G3abSec}

The symmetric component of the spectrum is given by
\begin{equation}
I = \{(k,l,m)| k,l,m \in \Z, l \geq 1, m = 0, \ldots, l-1 \},
\end{equation}
\begin{equation}
\lambda_{klm}^{\pm} = \pm 2 \pi \sqrt{\frac{1}{H^2}(k+c)^2+ \frac{1}{L^2}l^2 + \frac{1}{3L^2}(l-2m)^2 },
\end{equation}
with $c=1/2$ for the spin structure $(a)$ and $c=0$ for the spin structure $(b)$.  

The manifold $G3$ is unusual in that the multiplicity of $\lambda_{klm}^{\pm}$ is equal to twice the number of elements in $I$ which map to it.

\begin{thm}\label{thmG3ab}
On the manifold $G3$ with an arbitrary choice of spin structure, the non-perturbative
spectral action is given by
\begin{equation}\label{G3abSA}
\Tr (f(D^2/\Lambda^2) )= \frac{1}{\sqrt{3}}\left(\frac{\Lambda}{2 \pi} \right) ^3 HL^2\int_{\R ^3} f\left(u^2 + v^2 + t^2 \right)dudvdt  
\end{equation}
plus possible terms of order $O(\Lambda^{-\infty})$.
\end{thm}

\proof Notice that $\lambda_{klm}^{\pm}$ is invariant under the linear transformations $R,S,T$, given by
\begin{align*}
R(l) &= -l \\
R(m) &= -m
\end{align*}
\begin{align*}
S(l) &= m \\
S(m) &= l
\end{align*}
\begin{align*}
T(l) &= l - m \\
T(m) &= -m
\end{align*}

Let $\tilde{I} =  \{(k,l,m)| k,l,m \in \Z, l \geq 2, m = 1, \ldots, l-1 \}$.

\begin{figure}
\includegraphics[scale=0.4]{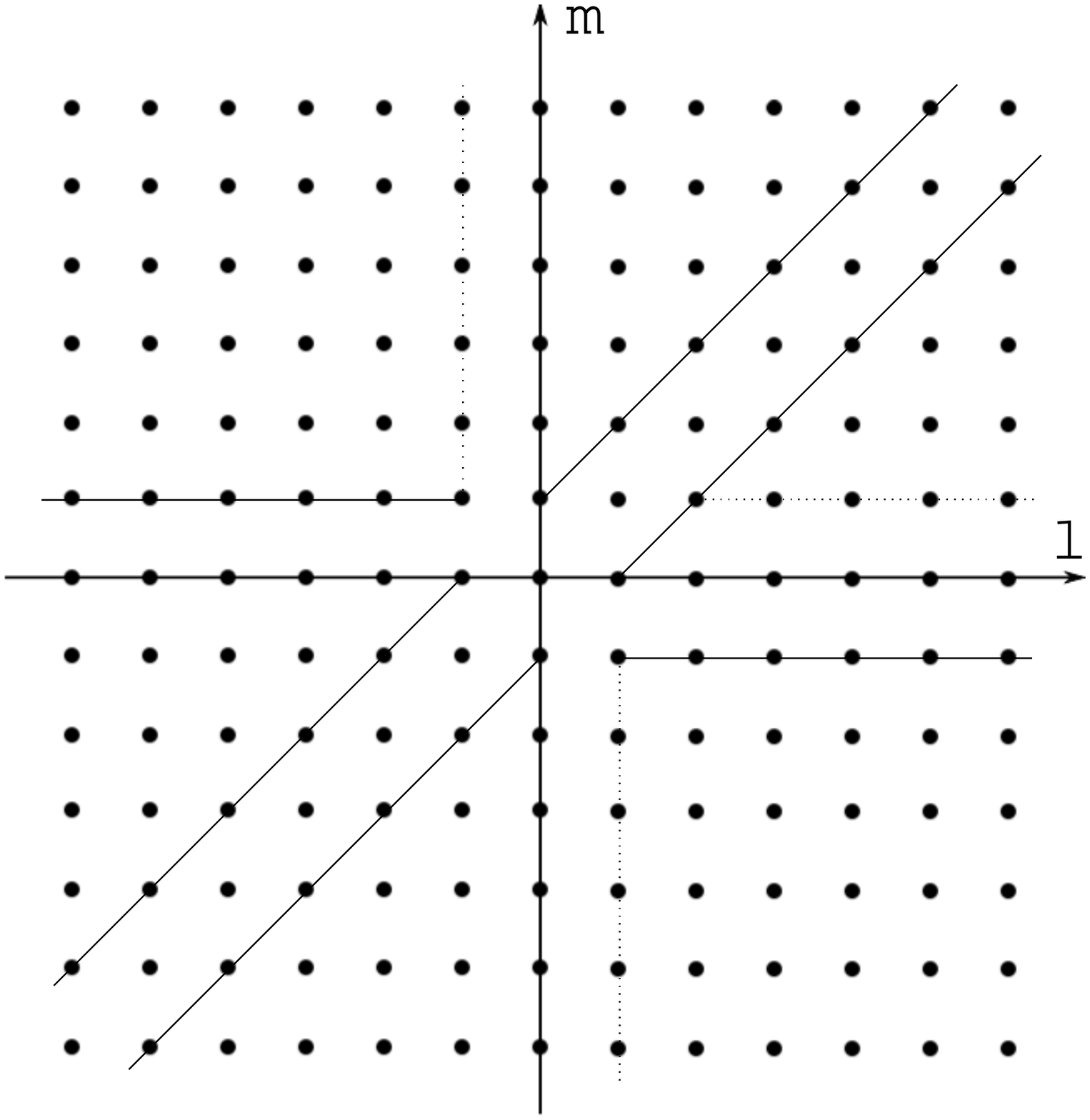}
\caption{Lattice decomposition for $G3$ computation.   Six regions and the set $l=m$.  The dashed lines indicate one of the boundary lines which define the region $\tilde{I}$ along with its images under the symmetries of $\lambda_{klm}$. The other boundary line of $\tilde{I}$ overlaps with the boundary of $I$. \label{g3Fig}}
\end{figure}

Then we may decompose $\Z ^3$ as (see figure \ref{g3Fig})
\begin{equation}
\Z ^3 = I \sqcup R(I) \sqcup S(I) \sqcup RS(I) \sqcup T(\tilde{I}) \sqcup RT (\tilde{I}) \sqcup \{l=m\}.
\end{equation}
Therefore, we have
\begin{align*}
\sum_{\Z ^3} f(\lambda_{klm}^2/\Lambda^2) &= 4\sum_I f(\lambda_{klm}^2/\Lambda^2) \\
& + 2\left( \sum_I f(\lambda_{klm}^2/\Lambda^2) - \sum_{m=0,~l\geq 1}f(\lambda_{klm}^2/\Lambda^2)  \right) \\ & + \sum_{l=m}f(\lambda_{klm}^2/\Lambda^2) \\
&= 6 \sum_I f(\lambda_{klm}^2/\Lambda^2) -  \sum_{m=0}f(\lambda_{klm}^2/\Lambda^2) \\ & + \sum_{m=0,~l=0}f(\lambda_{klm}^2/\Lambda^2) + \sum_{l=m}f(\lambda_{klm}^2/\Lambda^2) \\
\sum_I f(\lambda_{klm}^2/\Lambda^2) &= \frac{1}{6} \left(\sum_{\Z ^3} f(\lambda_{klm}^2/\Lambda^2)  + \sum_{m=0}f(\lambda_{klm}^2/\Lambda^2) \right) \\ &
- \frac{1}{6} \left(\sum_{m=0,~l=0}f(\lambda_{klm}^2/\Lambda^2) - \sum_{l=m}f(\lambda_{klm}^2/\Lambda^2) \right)
\end{align*}

Therefore the symmetric component of the spectrum contributes to the spectral action
\begin{align*}
& \frac{4}{6} ( \left( \frac{\Lambda}{2 \pi} \right) ^3H L^2 \int_{\R ^3} f(u^2 + v^2 + \frac{1}{3}(v-2w)^2) \\ & +  \left( \frac{\Lambda}{2\pi} \right) ^2 HL \int_{\R ^2} f(u^2 + \frac{4}{3}v^2)  - \left( \frac{\Lambda}{2\pi} \right)H  \int_{\R} f(u^2) \\ 
&-\left( \frac{\Lambda}{2\pi} \right) ^2 HL \int_{\R ^2} f(u^2 + \frac{4}{3}v^2) ) + O(\Lambda^{-\infty})  \\
&=  \frac{4}{6} \left( \frac{\Lambda}{2 \pi}^3 HL^2 \int_{\R ^3} f(u^2 + v^2 + \frac{1}{3}(v-2w)^2)  - \frac{\Lambda}{2\pi} H \int_{\R} f(u^2)  \right) \\ & + O(\Lambda^{-\infty})
\end{align*}

Combining this with the asymmetric contribution \eqref{asym},  we see that the spectral action of spaces $G3(a)$ and $G3(b)$ is equal to
\[
\frac{2}{3}\left(\frac{\Lambda}{2 \pi} \right) ^3 HL^2\int_{\R ^3} f\left(u^2 + v^2 + \frac{1}{3}(v-2w)^2 \right)dudvdw  + O(\Lambda^{-\infty}).
\]

Now, if one makes the change of variables $(u,v,w) \mapsto (u,v,t)$,  where
\[
t =  \frac{2w-v}{\sqrt{3}},
\]
then the spectral action becomes
\[
\frac{1}{\sqrt{3}}\left(\frac{\Lambda}{2 \pi} \right) ^3 HL^2\int_{\R ^3} f\left(u^2 + v^2 + t^2 \right)dudvdt  + O(\Lambda^{-\infty}).
\]
\endproof

Notice that, a priori, one might have expected a possibly different result in this case,
because the Bieberbach manifold is obtained starting from a hexagonal lattice rather
than the square lattice, but up to a simple change of variables in the integral, this gives
again the same result, up to a multiplicative constant, as in the case of the standard 
flat torus.

\subsection{The spectral action for $G4$}\label{G4sec}

The Bieberbach manifold $G4$ is referred to in \cite{RWULL}  as the ``quarter-turn space". It is obtained by considering a lattice generated by the vectors $a_1=(0,0,H)$, $a_2=(L,0,0)$, and
$a_3=(0,L,0)$, with $H,L>0$, and taking the quotient of $\R^3$ by the group $G4$ generated by the commuting translations $t_i$ along the vectors $a_i$ and an additional generator $\alpha$
with the relations
\begin{equation}\label{refG4}
\alpha^4 = t_1, \ \ \  \alpha t_2 \alpha^{-1} = t_3, \ \ \  \alpha t_3 \alpha^{-1} = t_2^{-1}.
\end{equation}
This produces an identification of the sides of a fundamental domain with a rotation
by an angle of $\pi/2$ about the $z$-axis.
Theorem 3.3 of \cite{Pfa} shows that the manifold $G4$ has four different spin structures
parameterized by two signs $\delta_i =\pm 1$. There are correspondingly two different forms 
of the Dirac spectrum, as shown in Theorem 5.7 of \cite{Pfa}, one for $\delta_i=\{ \pm 1, 1\}$,
the other for $\delta_i=\{ \pm 1, -1 \}$, denoted by $G4(a)$ and $G(4)b$. 

Again the nonperturbative spectral action is independent of the spin structure and equal in
both cases to the same multiple of the spectral action for the torus.

\subsubsection{The case of $G4(a)$}\label{G4aSec}

\begin{thm}\label{thmG4a}
On the manifold $G4$ with a spin structure $(a)$ with $\delta_i=\{ \pm 1, 1\}$, the non-perturbative
spectral action is given by
\begin{equation}\label{G4aSA}
\Tr (f(D^2/\Lambda^2) )= \frac{1}{2} \left( \frac{\Lambda}{2\pi} \right)^3 HL^2 \int_{\R ^3}f(u^2 + v^2 + w^2)du dv dw
\end{equation}
plus possible terms of order $O(\Lambda^{-\infty})$.
\end{thm}

\begin{figure}
\includegraphics[scale=0.4]{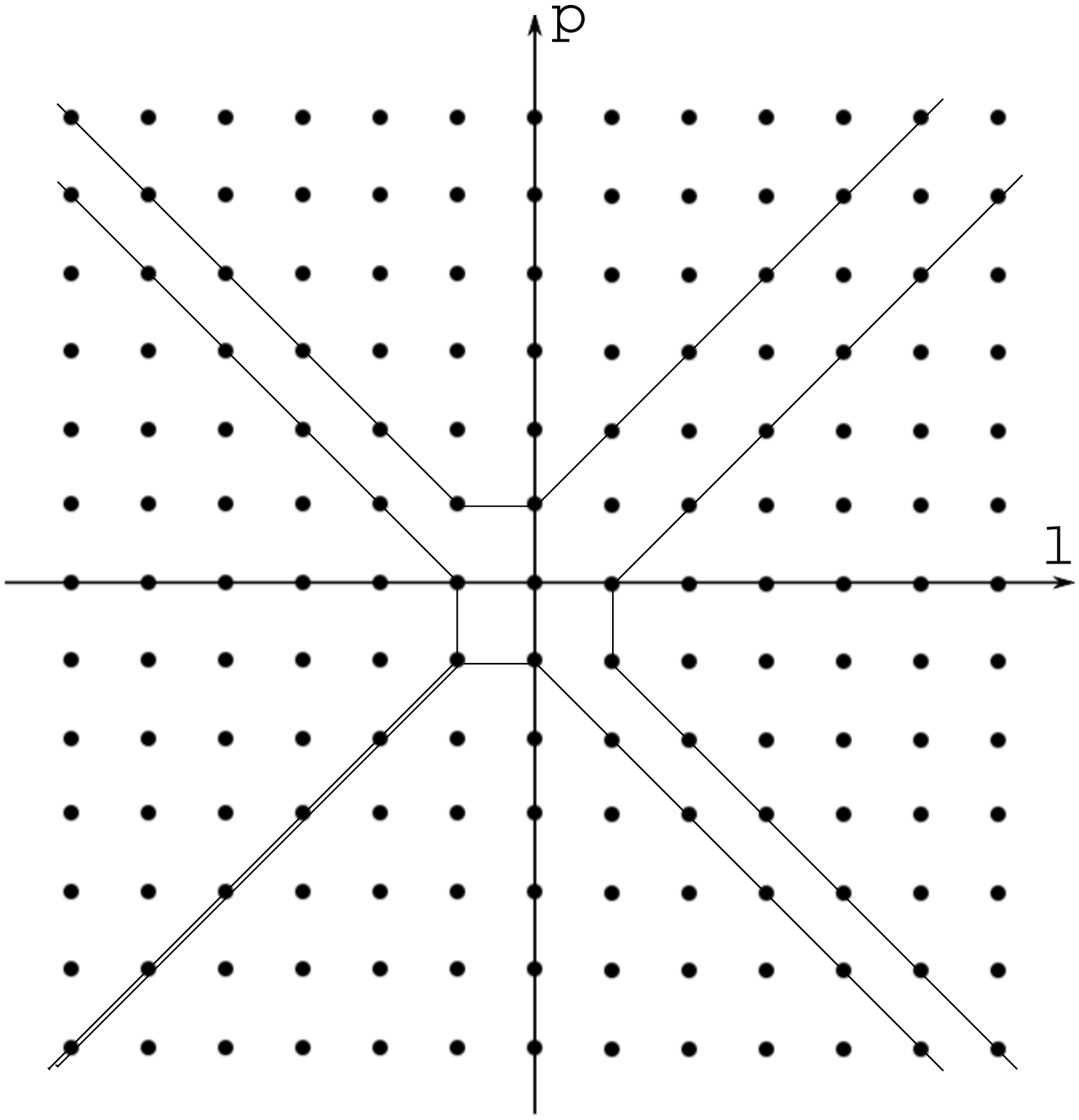}
\caption{Lattice decomposition for $G4(a)$ computation.  Four regions.   \label{g4aFig}}
\end{figure}

\proof The symmetric component of the spectrum is given by
\[
I = \{ (k,l,m) | k,l,m \in \Z, l \geq 1, m = 0, \ldots , 2l-1\}
\]
\[
\lambda_{klm}^{\pm} = \pm 2 \pi \sqrt{\frac{1}{H^2}(k+\frac{1}{2})^2+ \frac{1}{L^2}(l^2 + (m-l))^2  },
\]
First, we make the change of variables $p = m-l$.  Then we use the symmetries
\begin{align*}
l \mapsto -l \\
l \mapsto p \quad p\mapsto l \\
l \mapsto p \quad p \mapsto -l\\
\end{align*}
to cover all of $\Z ^3$ except for the one-dimensional lattice $\{(k,l,p)| l=p=0\}$.  This decomposition is depicted in figure \ref{g4aFig}.  In the figure one sees that the points $l = p$ such that $l < 0$ are covered twice, and the points $l=p$ such that $l > 0$ are not covered at all, but via the transformation $(l,p) \mapsto -(l,p)$, this is the same as covering each of the points $l = p$, $l \neq 0$ once.  Observations like this will be suppressed in the sequel.  Then we see that the contribution from the symmetric component of the spectrum to the spectral action is
\begin{equation}
\frac{1}{2} \left( \frac{\Lambda}{2\pi} \right)^3 HL^2 \int_{\R ^3}f(u^2 + v^2 + w^2)du dv dw - \frac{1}{2} \left( \frac{\Lambda}{2\pi} \right) H \int_{\R}f(u^2)du  ,
\end{equation}
up to terms of order $O(\Lambda^{-\infty})$.
Combining this with the asymmetric component, we find that the spectral action is given by
\eqref{G4aSA}.
\endproof

\subsubsection{The case of $G4(b)$}\label{G4bSec}

In this case there is no asymmetric component in the spectrum.  
The symmetric component is given by the data
\[
I = \{ (k,l,m) | k,l,m \in \Z, l \geq 1, m = 0, \ldots , 2l-2\}
\]
\[
\lambda_{klm}^{\pm} = \pm 2 \pi \sqrt{\frac{1}{H^2}(k+\frac{1}{2})^2+ \frac{1}{L^2}((l-1/2)^2 + (m-l + 1/2))^2  }.
\]
We again obtain the same expression as in the $G4(a)$ case for the
spectral action.

\begin{thm}\label{thmG4b}
On the manifold $G4$ with a spin structure $(b)$ with $\delta_i=\{ \pm 1, -1\}$, the non-perturbative
spectral action is also given by
\begin{equation}\label{G4bSA}
\Tr (f(D^2/\Lambda^2) )= \frac{1}{2} \left( \frac{\Lambda}{2\pi} \right)^3 HL^2 \int_{\R ^3}f(u^2 + v^2 + w^2)du dv dw
\end{equation}
up to possible terms of order $O(\Lambda^{-\infty})$.
\end{thm}

\begin{figure}
\includegraphics[scale=0.4]{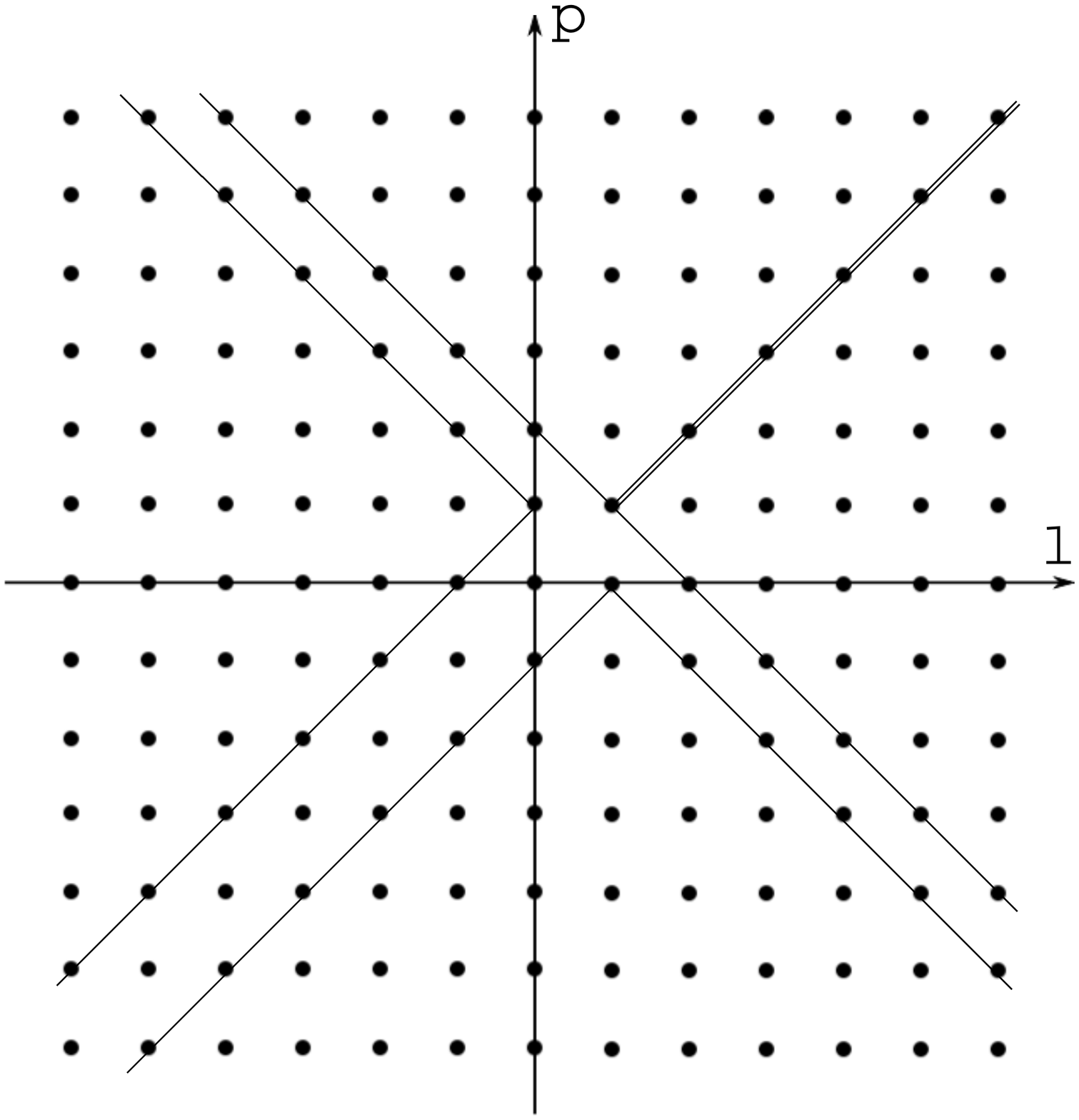}
\caption{Lattice decomposition for $G4(b)$ computation.  Four regions. \label{g4bFig}}
\end{figure}

\proof  We make the change of variables $p = m-l$.  Using the symmetries
\begin{align*}
l \mapsto 1-l \\
l \mapsto p \quad p\mapsto l \\
l \mapsto p \quad p \mapsto 1 - l,\\
\end{align*}
we can exactly cover all of $\Z ^3$, as shown in figure $\ref{g4bFig}$ and so the spectral action has the expression
\eqref{G4bSA}.
\endproof

\begin{rem}\label{noG5}{\rm
The technique we use here to sum over the spectrum to compute
the non-perturbative spectral action does not appear to work in the
case of the Bieberbach manifold $G5$, which is the ``sixth-turn space"
described from the cosmic topology point of view in \cite{RWULL},
namely the quotient of $\R^3$ by the group $G5$ generated by
commuting translations $t_i$ along the vectors $a_1=(0,0,H)$,
$a_2=(L,0,0)$ and $a_3=(\frac{1}{2}L, \frac{\sqrt{3}}{2}L,0)$, $H,L>0$,
and an additional generator $\alpha$ with $\alpha^6=t_1$, $\alpha t_2 \alpha^{-1}=t_3$
and $\alpha t_3 \alpha^{-1}= t_2^{-1} t_3$, which produces
an identification of the faces of the fundamental domain with
a $\pi/3$-turn about the $z$-axis. This case will therefore be analyzed elsewhere, but
it is reasonable to expect that it will also give a multiple of the spectral action of the torus,
with a proportionality factor of $HL^2/(4\sqrt{3})$.
}\end{rem}

\subsection{The spectral action for $G6$}\label{G6sec}
We analyze here the last remaining case of compact orientable
Bieberbach manifold $G6$, the Hantzsche--Wendt space, according to
the terminology followed in \cite{RWULL}. This is the quotient of $\R^3$
by the group $G6$ obtained as follows. One considers the lattice generated
by vectors $a_1=(0,0,H)$, $a_2=(L,0,0)$, and $a_3=(0,S,0)$, with $H,L,S>0$,
and the group generated by commuting translations $t_i$ along these vectors,
together with additional generators $\alpha$, $\beta$, and $\gamma$ with the
relations
\begin{equation}\label{relG6}
\begin{array}{ccc}
\alpha^2 =t_1, & \alpha t_2 \alpha^{-1} = t_2^{-1}, & \alpha t_3 \alpha^{-1} = t_3^{-1}, \\
\beta^2 = t_2, & \beta t_1 \beta^{-1} =t_1^{-1}, & \beta t_3 \beta^{-1} = t_3^{-1}, \\
\gamma^2 = t_3, & \gamma t_1 \gamma^{-1} = t_1^{-1}, & \gamma t_2 \gamma^{-1} = t_2^{-1}, \\
& \gamma\beta\alpha = t_1 t_3.  & 
\end{array}
\end{equation}
This gives an identification of the faces of the fundamental domain with a
twist by an angle of $\pi$ along each of the three coordinate axes. 

According to Theorems 3.3 and 5.7 of \cite{Pfa}, the manifold $G6$ has four different
spin structures parameterized by three signs $\delta_i=\pm$ subject to the constraint
$\delta_1\delta_2\delta_3=1$, but all of them yield the same Dirac spectrum, which has
the following form.

The manifold $G6$ also has no asymmetric component to its spectrum, while
the symmetric component is given by
\[
I = \{ (k,l,m) | k,l,m \in \Z, l \geq 0, k \geq 0\}
\]
\[
\lambda_{klm}^{\pm} = \pm 2 \pi \sqrt{\frac{1}{H^2}(k+\frac{1}{2})^2+ \frac{1}{L^2}(l+\frac{1}{2})^2 + \frac{1}{S^2}(m + \frac{1}{2})^2 }.
\]
We then obtain the following result.

\begin{thm}\label{thmG6}
The Bieberbach manifold $G6$ with an arbitrary choice of spin structure has 
nonperturbative spectral action of the form
\begin{equation}\label{G6SA}
\Tr f(D^2 / \Lambda^2) = \frac{1}{2} \left( \frac{\Lambda}{2\pi} \right)^3 HLS \int_{\R ^3}f(u^2 + v^2 + w^2)du dv dw 
\end{equation}
up to terms of order $ O(\Lambda^{-\infty})$.
\end{thm}

\begin{figure}
\includegraphics[scale=0.4]{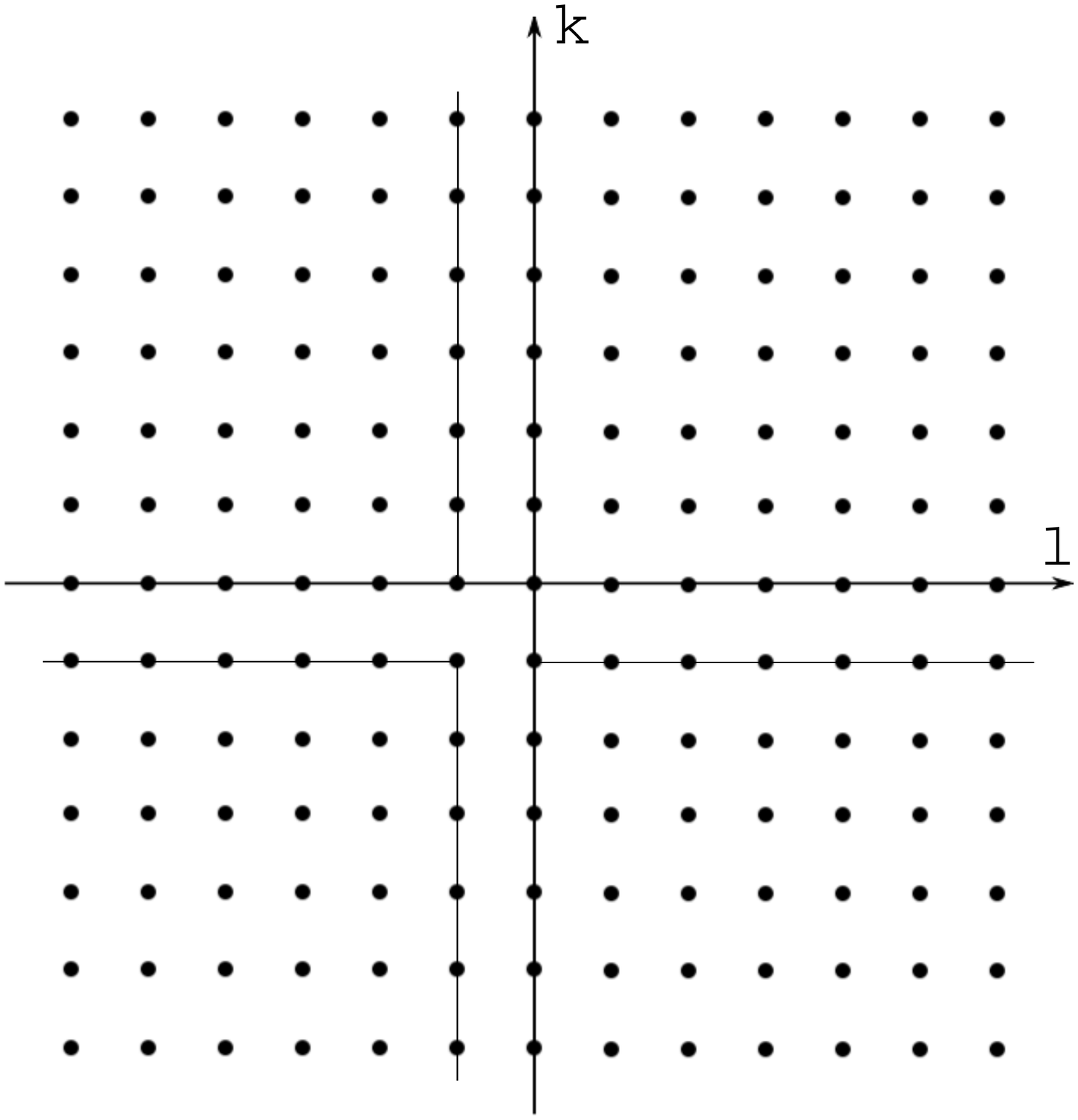}
\caption{Lattice decomposition for $G6$ computation.  Four regions.  \label{g6Fig}}
\end{figure}

\proof
Using the three transformations 
\begin{align*}
k \mapsto -k -1, \\
l \mapsto -l -1,\\
k\mapsto -k-1 \quad l\mapsto -l-1,
\end{align*}
one exactly covers $\Z ^3$, as seen in figure \ref{g6Fig} and so we see that the nonperturbative spectral action is given by
\eqref{G6SA}.
\endproof

\section{Geometry, topology and inflation: flat manifolds}\label{FlatPowerSec}

As shown in Theorem 8.3 of \cite{MaPieTeh}, on a flat torus of sides $\ell=1$ the
slow roll potential is of the form
$$ V_{T^3}(\phi)= \frac{\Lambda^4 \beta}{4\pi} \cV_{T^3}(\phi^2/\Lambda^2), $$
with $\cV_{T^3}(x)$ given by
\begin{equation}\label{VT3}
\cV_{T^3}(x)= \int_0^\infty u \, (h(u+x) - h(u))\, du ,
\end{equation} 
as in \eqref{VWS3}.

\begin{prop}\label{T3potential}
Let $Y$ be a Bieberbach manifold $Y=T^3/\Gamma$ with the induced flat metric. 
Then the slow-roll potential $V_Y(\phi)$ in \eqref{Spact4Dphi}
is of the form
\begin{equation}\label{VphiT3Y}
V_Y(\phi) =  \frac{\Lambda^4 \beta}{4\pi} \cV_Y (\frac{\phi^2}{\Lambda^2}),
\end{equation}
where 
\begin{equation}\label{VYT3}
\cV_Y(x) = \lambda_Y \, \cV_{T^3}(x) 
\end{equation}
with $\cV_{T^3}(x)$ as in \eqref{VT3} and the factor $\lambda_Y$ given by 
\begin{equation}\label{lambdaYflat}
\lambda_Y  = \left\{ \begin{array}{ll}
\frac{HSL}{2} & \Gamma=G2 \\[3mm]
\frac{HL^2}{2\sqrt{3}} & \Gamma =G3 \\[3mm]
\frac{HL^2}{4} & \Gamma =G4 \\[3mm]
\frac{HLS}{4} & \Gamma=G6
\end{array}\right.
\end{equation}
\end{prop}

\proof The result follows directly from the theorems proved in \S \ref{BiebSpSec}
above, which show that the nonperturbative spectral action for $Y$ is a multiple
of the spectral action for $T^3$ with proportionality factor given by $\lambda_Y$
as in \eqref{lambdaYflat}. The potentials
$$ \Tr(h((D^2_{Y\times S^1}+\phi^2)/\Lambda^2)) - \Tr(h(D^2_{Y\times S^1}/\Lambda^2)) = V_Y(\phi) $$
are then related by the same proportionality factor $\lambda_Y$.
\endproof

We the obtain the following analog of Proposition \ref{PowerSpheres} in the flat case.

\begin{prop}\label{PowerFlat}
Let $\cP_{s,Y}(k)$ and $\cP_{t,Y}(k)$ denote the power spectra for the density
fluctuations and the gravitational waves, computed as in \eqref{PstV}, for the 
slow-roll potential $V_Y(\phi)$. Then they satisfy the power law
\begin{equation}\label{powerlawPY}
\begin{array}{rl}
\cP_{s,Y}(k) \sim & \lambda_Y \, \cP_s(k_0) 
\displaystyle{\left(\frac{k}{k_0} \right)^{1 - n_{s,T^3} + \frac{\alpha_{s,T^3}}{2} \log(k/k_0)}}
\\[3mm]  \cP_{t,Y}(k) \sim & \lambda_Y \, \cP_t(k_0) \displaystyle{\left(\frac{k}{k_0} 
\right)^{n_{t,T^3} + \frac{\alpha_{t,T^3}}{2} \log(k/k_0)}} , \end{array}
\end{equation}
where $\lambda_Y$ is as in \eqref{lambdaYflat}, for $Y=T^3/\Gamma$ a Bieberbach manifold
and the spectral parameters
$n_{s,T^3}$, $n_{t,T^3}$, $\alpha_{s,T^3}$, $\alpha_{t,T^3}$ are computed as in
\eqref{spectralparam} from the slow-roll parameters \eqref{slowrollparam}, which satisfy
$\epsilon_Y = \epsilon_{T^3}$, $\eta_Y=\eta_{T^3}$, $\xi_Y=\xi_{T^3}$.
\end{prop}

\begin{figure}
\includegraphics[scale=0.8]{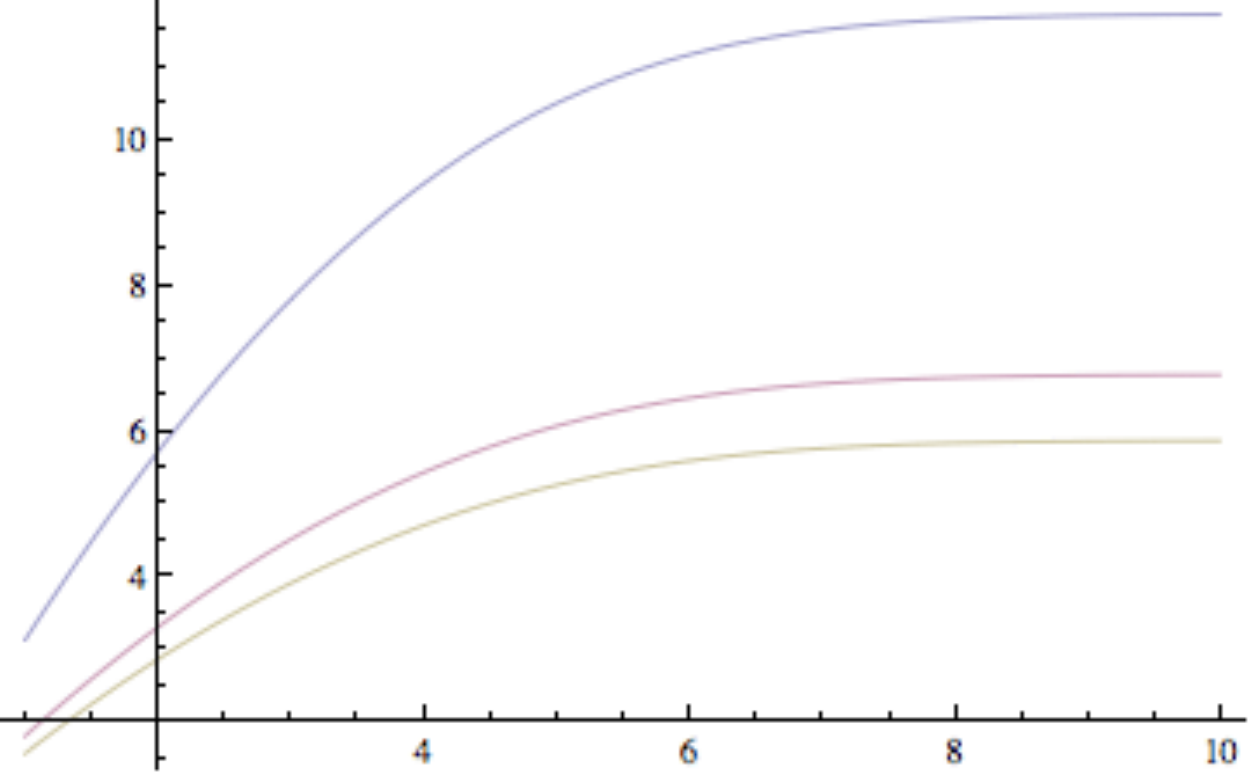}
\caption{The slow-roll potentials for the $G2$ case (top curve), the $G3$ case (middle curve), and the $G4$ and $G6$ cases (lower curve). \label{VflatFig}}
\end{figure}

If one assumes that each of the characteristic sizes involved, $H$, $L$, $S$ would
be comparable to $\Lambda^{-1}$, after Wick rotating from Euclidean to
Lorentzian signature, as in the expansion scale $\Lambda(t) \sim 1/a(t)$, one
would then obtain proportionality factors that are simply of the form
\begin{equation}\label{lambdaYflat2}
\lambda_Y  = \left\{ \begin{array}{ll}
\frac{1}{2} & \Gamma=G2 \\[3mm]
\frac{1}{2\sqrt{3}} & \Gamma =G3 \\[3mm]
\frac{1}{4} & \Gamma =G4 \\[3mm]
\frac{1}{4} & \Gamma=G6
\end{array}\right.
\end{equation}
Assuming then that $\Lambda\beta=1$, and 
using the same test function $h_n(x)$ with $n=20$ as in Figure \ref{hnFig} we then
obtain different curves as in Figure \ref{VflatFig} for the $G2$ case (top curve), $G3$ case
(middle curve), and for the $G4$ and $G6$ cases (bottom curve).

\section{Conclusions: Inflation potential, power spectra, and cosmic topologies}\label{InflPotSec}

We have seen in this paper that, in a modified gravity model based on the non-perturbative
spectral action functional, different cosmic topologies, either given by spherical space forms
or by flat Bieberbach manifolds, leave a signature that can distinguish between the different
topologies in the form of the slow roll inflation potential that is obtained from the variation of
the spectral action functional. The amplitude of the potential, and therefore the amplitude of
the corresponding power spectra for density perturbations and gravitational waves (scalar and tensor perturbations), differs by a factor that depends on the topology, while the slow-roll
parameters only detect a difference between the spherical and flat cases. As one knows from
\cite{Lidsey},  \cite{SKamCoo}, \cite{StLy}, both the slow-roll parameters and the amplitude
of the power spectra are constrained by cosmological information, so in this kind of 
modified gravity model, one in principle obtains a way to constrain the topology of the
universe based on the slow-roll inflation potential, on the slow-roll parameters and
on the power spectra for density perturbations and gravitational waves. The factors 
$\lambda_Y$ that correct the amplitudes depending on the topology are given by the
following table.

\begin{center}
\begin{tabular}{| c | c || c | c |}
\hline
$Y$ spherical & $\lambda_Y$ &
$Y$ flat & $\lambda_Y$ \\ \hline
& & & \\
sphere & $1$ & flat torus & $1$ \\
& & & \\
\hline
& & & \\
lens $N$ &  $\frac{1}{N}$ & $G2(a)(b)(c)(d)$ & $\frac{HSL}{2}$  \\ 
& & & \\
 \hline
 & & & \\
binary dihedral $4N$ & $\frac{1}{4N}$ & $G3(a)(b)$ & $\frac{HL^2}{2\sqrt{3}}$ \\ 
& & & \\
\hline
& & & \\
binary tetrahedral & $\frac{1}{24}$ & $G4(a)(b)$ & $\frac{HL^2}{4}$ \\
& & & \\
 \hline
& & & \\
binary octahedral & $\frac{1}{48}$ & $G5$ & ? \\  
& & & \\
\hline
& & & \\
binary icosahedral & $\frac{1}{120}$ & $G6$ &  $\frac{HLS}{4}$ \\ 
& & & \\
\hline
\end{tabular}
\end{center}

\medskip

Notice that some ambiguities remain: the form of the potential and the value of
the scale factor $\lambda$ alone do not distinguish, for instance, between a
lens space with $N=24$, a binary dihedral quotient with $N=6$ and the
binary tetrahedral quotient, or between the Poincar\'e dodecahedral space (the
binary icosahedral quotient), a lens space of order $N=120$ and a binary dihedral
quotient with $N=30$. At this point we do not know whether more refined
information can be extracted from the spectral action that can further
distinguish between these cases, but we expect that, when taking into account
a more sophisticated version of the spectral action model, where gravity is
coupled to matter by the presence of additional (non-commutative) small
extra-dimensions (as in \cite{CCM}, \cite{CoSM}), one may be able to
distinguish further. In fact, instead of a trivial product $X\times F$, one
can include the non-commutative space $F$ using a topologically
non-trivial fibration over the 4-dimensional spacetime $X$ and this allows
for a more refines range of proportionality factors $\lambda_Y$. We will
discuss this in another paper.

\end{document}